\documentclass[proof]{pasj00}
\usepackage{graphicx}
\Received{}
\Accepted{2015 January 13}
\Published{$\langle$publication date$\rangle$}
\SetRunningHead{Astronomical Society of Japan}{Usage of \texttt{pasj00.cls}}
\def \vec#1{\mbox{\boldmath $#1$}} %ベクトルマクロ
\def \fourier#1{\mathcal{F}(#1)}   %フーリエ演算
\usepackage{times}

\begin{document}

\title{Smoothed Particle Hydrodynamics with Smoothed Pseudo-Density}
%my comment %if no document option isn't proof it wouldn't be needed 2 line. 

\author{Satoko \textsc{Yamamoto},\altaffilmark{1}
        Takayuki R. \textsc{Saitoh},\altaffilmark{2}
        and
        Junichiro \textsc{Makino}\altaffilmark{2,3}
        }
\altaffiltext{1}{Department of Earth \& Planetary Science, Tokyo Institute of Technology,\\
   Ookayama, Meguroku, Tokyo 152-8551}
\email{yamamoto.s.an@geo.titech.ac.jp}
\altaffiltext{2}{Earth-Life Science Institute, Tokyo Institute of Technology,\\
   2-12-1 Ookayama, Meguro, Tokyo 152-8550}
\altaffiltext{3}{RIKEN Advanced Institute for Computational Science, Minatojima-minamimachi,\\
   Chuo-ku, Kobe, Hyogo 650-0047}


\maketitle

\begin{abstract}
In this paper, we present a new formulation of smoothed particle hydrodynamics (SPH), 
which, unlike the standard SPH (SSPH), is well-behaved at the contact discontinuity. 
The SSPH scheme cannot handle discontinuities in density
 (e.g. the contact discontinuity and the free surface), 
because it requires that the density of fluid is positive and continuous everywhere.  
Thus there is inconsistency in the formulation of the SSPH scheme at discontinuities 
of the fluid density. 
To solve this problem, we introduce a new quantity associated with particles 
and ``density'' of that quantity. 
This ``density'' evolves through the usual continuity equation 
with an additional artificial diffusion term, in order to guarantee the continuity 
of ``density''. 
We use this ``density'' or pseudo density, instead of the mass density,
to formulate our SPH scheme. 
We call our new method as SPH with smoothed pseudo-density (SPSPH). 
We show that our new scheme is physically consistent and can handle discontinuities 
quite well. 
\end{abstract}

%\section{Overview}
%When \texttt{pasj00.cls} is applied to an article for PASJ, 
%the article should be prepared in the standard \LaTeXe{} style with 
%slight modifications.  
%That is, a manu\-script has the following structure:
%\begin{verbatim}
%\documentclass{pasj00}

%\begin{document}

%\author{list of authors}
%\affil{the authors' affiliation}
%\title{title of the article}
%%% some other commands

%\maketitle
%
%\begin{abstract}
%  abstract of the article
%\end{abstract}

%イントロ%%%%%%%%%%%%%%%%%%%%%%%%%%%%%%%%%%%%%%%%%%%%%%%%%%%%%%%%%%%%%%%%%%%%%%%%%%%%%%%%%%%
\section{Introduction}
Smoothed particle hydrodynamics (SPH) is one of the methods to solve 
the equations of fluid by expressing fluid as a collection of fluid particles. 
It was proposed by Lucy (1977) and Gingold \& Monaghan (1977). 
It is suitable for systems with large voids or 
systems which exhibit large structural changes. 
Thus it has been widely used for simulations of planetary science and astrophysics.

%%%SSPH問題点%%%
Recently, however, it has been reported that standard SPH (SSPH) suppresses the Kelvin-Helmholtz instability (Here after KHI) (Agertz et al. 2007, Okamoto et al. 2003). 
The reason for this suppression is that SSPH requires that density is positive and continuous even at the discontinuities of density (e.g. the contact discontinuity). 
This inconsistency causes large errors in the pressure estimate, 
which then causes unphysical repulsive force between particles 
in the low- and high-density region.

%%%先行研究%%%
So far, a number of solutions have been proposed for this problem. 
Examples are the use of pressure (energy density) instead of the mass density for the SPH formulation 
(we call this formulation density independent formulation of SPH, DISPH)
(Ritchie \& Thomas 2001, Read et al. 2010, Saitoh \& Makino 2013, Hopkins 2013, Hosono et al. 2013, Rosswog 2014).
Price (2008) introduced an artificial thermal conductivity (AC) which the pressure distribution smooths.
Cha et al. (2010) and Murante et al. (2011) employed Godunov SPH (GSPH) proposed by Inutsuka (2002). 
Each of these solutions still has its intrinsic disadvantages. 
For instance, DISPH has difficulties when the pressure is close to zero, as is the case at the free surface,
while the AC term does not exist in the original Euler equation. We will revisit this point in section 2.2.

%%%新しいSPH%%%
In this paper, we propose a new formulation of SPH which can in principle handle systems with any discontinuity. 
We introduce a new quantity $y$ which we call pseudo-density, and require the continuity and positivity of $y$ instead of that of the density. 
The quantity $y$ follows the continuity equation like that of the density, 
but with an additional diffusion term. 
We can set arbitrarily value as the initial condition for $y$ as long as it is positive.
Except for the initial moment, $y$ is always continuous and positive, as the result of diffusion, 
even if we give discontinuous initial distribution. 
Note that the introduction of $y$, which is the density of something, 
means we introduced effectively an extensive quantity associated to particle. 
We use the symbol $Z$ for this quantity and we call it pseudo-mass. 
This $y$-$Z$ pair is used only to construct the SPH approximation. 
Though they diffuse following the diffusion equation, 
this diffusion does not introduce numerical error beyond the discretization 
error of the SPH scheme, even at discontinuities. 
We call this formulation of SPH, smoothed pseudo-density SPH (SPSPH). 
We will show that for contact discontinuities in general SPSPH gives the results better than or at least as good as that of SSPH. If we choose the initial pseudo-density adequately, SPSPH gave the result better than that of SSPH for all tests we tried so far.
Moreover it might be extended to handle free surface.

%%%論文流れ%%%
The structure of this paper is as follows. 
In section 2, we describe the problem of SSPH and solutions proposed so far. 
We also discuss problems that have not been solved in previous studies. 
We propose our new method, SPSPH, in section 3. 
In section 4 we present the comparison of the results of test calculations with SPSPH and SSPH. Finally we present discussion in section 5 and summary in section 6.

%%%%%%%%%%%%%%%%%%%%%%%%%%%%%%%%%%%%%%%%%%%%%%%%%%%%%%%%%%%%%%%%%%%%%%%%%%%%%%%%%%%%%%%%%%%%
%%% contents

\section{Standard SPH and Its problem}

%SPH一般的定式化%%%%%%%%%%%%%%%%%%%%%%%%%%%%%%%%%%%%%%%%%%%%%%%%%%%%%%%%%%%%%%%%%%%%%%%%%%%%

\subsection{Formulation for the standard SPH}
In SPH, the fluid is expressed by a collection of fluid particles. Physical quantities of fluid particles are approximated by the convolution of the quantity with the kernel function. 
The kernel convolution of a physical quantity $f$ at the position $\vec{r}$ is defined 
\begin{eqnarray}
\langle f \rangle(\vec{r}) = \int^\infty_{-\infty} f(\vec{r}')W(|\vec{r}'-\vec{r}|,h) d\vec{r'}.
\end{eqnarray}

The function $W(|\vec{r}'-\vec{r}|,h)$ is the kernel function and $h$ is the smoothing length. The kernel function must satisfy the following four properties: 
(I) it must converge to the $\delta$ function in the limit of $h\rightarrow $ 0,
(I\hspace{-.1em}I) its integration is normalized to unity, 
(I\hspace{-.1em}I\hspace{-.1em}I) it is a function with a compact support, 
(I\hspace{-.1em}V) it can be differentiated at least once $\left[C^{1}(\vec{r}){\rm class}\right]$. 

The first order derivative, $\vec{\nabla}_{\vec{r}}f(\vec{r})$, is given by
\begin{eqnarray}
\langle\vec{\nabla}_{\vec{r}} f\rangle(\vec{r}) = \int^\infty_{-\infty} \vec{\nabla}_{\vec{r}'}f(\vec{r}')W(|\vec{r}'-\vec{r}|,h) d\vec{r'}=\int^\infty_{-\infty} f(\vec{r}')\vec{\nabla}_{\vec{r}}W(|\vec{r}'-\vec{r}|,h) d\vec{r'},
\end{eqnarray}
by applying the partial integral and using the property (I\hspace{-.1em}I\hspace{-.1em}I) of the kernel function.
%
%離散化開始%%%%%%%%%%%%%%%%%%%%%%%%%%%%%%%%%%%%%%%%%%%%%%%%%%%%%%%%%%%%%%%%%%%%%%%%%%%%%%%%%
%
In order to evaluate the value of a physical quantity $f(\vec{r})$ at the position of a fluid particle, we discretize Eq.(1) as follows,
\begin{eqnarray}
\langle f_a\rangle(\vec{r}_a) = \sum_b f_bW(|\vec{r}_{ab}|,h_a) \Delta V_b,
\end{eqnarray}
where the subscripts $a$ and $b$ denote particle  indices, $f_b$ is the value of $f$ at the position of particle $b$ and $\vec{r}_{ab} = \vec{r}_a-\vec{r}_b$.
In SSPH, $\Delta V_b$ is replaced by $m_b/\rho_b$, 
where $m_b$ and $\rho_b$ are the mass and the density of particle $b$. 
Thus $f_a$ is given by
\begin{eqnarray}
f_a = \sum_b f_b\frac{m_b}{\rho_b}W_{ab}(h_a),
\end{eqnarray}
where $W_{ab}(h_a)=W(|\vec{r}_a-\vec{r}_b|,h_a)$. 
From Eq.(2), $\vec{\nabla}_af_a$ is given by
\begin{eqnarray}
\vec{\nabla}_af_a = \sum_b f_b\frac{m_b}{\rho_b}\vec{\nabla}_aW_{ab}(h_a),
\end{eqnarray}
where the operator $\vec{\nabla}_a$ denotes the first order differentiation by $\vec{r}_a$.
In this paper, we determine $h_a$ by the following equation:
\begin{eqnarray}
h_a = \eta\left(\frac{1}{n_a}\right)^{1/D}.
\end{eqnarray}
Here, $n_a$ is the number density of particles at the position of particle $a$, 
and $D$ is the number of dimensions. 
The number density $n_a$ is defined as 
\begin{eqnarray}
n_a \equiv \sum_b W_{ab}(h_a).
\end{eqnarray}
The coefficient $\eta$ is a positive constant and equal to 1.6 in this paper. 

%SSPH定式化%%%%%%%%%%%%%%%%%%%%%%%%%%%%%%%%%%%%%%%%%%%%%%%%%%%%%%%%%%%%%%%%%%%%%%%%%%%%%%%%%

Now we derive the expression of fundamental equations of fluid for the standard SPH (SSPH).
In SSPH the density $\rho$ is obtained by substituting $\rho$ for $f$ in Eq.(4),
\begin{eqnarray}
\rho_a = \sum_b m_bW_{ab}(h_a).
\end{eqnarray}

%%

%運動方程式定式化開始%%%%%%%%%%%%%%%%%%%%%%%%%%%%%%%%%%%%%%%%%%%%%%%%%%%%%%%%%%%%%%%%%%%%%%%%

%ラグランジアン準備%%%%%%%%%%%%%%%

First we derive the equation of motion from the SPH Lagrangian (Springel \& Hernquist 2002; Rosswog 2009; Springel 2010; Hopkins 2013). The Lagrangian is given by
\begin{eqnarray}
L(\vec{q}) = \sum_b \frac{1}{2}(m_b \vec{v}_b^2)-\sum_bm_bu_b + \sum_b \lambda_b\phi_b,
\end{eqnarray}
where
\begin{eqnarray}
\vec{q} &=& (\vec{r}_1,\vec{r}_2,\cdots,\vec{r}_b,\cdots,h_1,h_2,\cdots,h_b,\cdots),\\
u_b &=& \frac{A_b}{\gamma-1} \rho_b^{\gamma-1}.
\end{eqnarray}
The quantity $A$ depends only on the entropy and is defined by $A\equiv \rho^\gamma/P$, and $\phi$ is the constraint for the smoothing length and equal to
\begin{eqnarray}
\phi_b = h_b - \eta\left(\frac{1}{n_b}\right)^{1/D}.
\end{eqnarray}

%%%%%%%%%%%%%%%%%%%%%%%%%%%%%%%%%%

%h空間での方程式変形%%%%%%%%%%%%%%%
By solving the Euler-Lagrange equation for $h_a$, we obtain
\begin{eqnarray}
\lambda_a = \frac{P_am_a}{\rho_a^2}\frac{\partial \rho_a}{\partial h_a}\left(1+\frac{h_a}{Dn_a}\frac{\partial n_a}{\partial h_a}\right)^{-1},
\end{eqnarray}
where 
\begin{eqnarray}
\frac{\partial \rho_a}{\partial h_a}&=&\sum_bm_b\partial_{h_a}W_{ab}(h_a),\\
\frac{\partial n_a}{\partial h_a}&=&\sum_b\partial_{h_a}W_{ab}(h_a),
\end{eqnarray}
and $\partial_{h_a}$ is $\partial/\partial h_a$.
%%%%%%%%%%%%%%%%%%%%%%%%%%%%%%%%%%

%r空間での方程式変形%%%%%%%%%%%%%%%

Now, we can solve the Euler-Lagrange equation for $\vec{r}_a$, which is given by
\begin{eqnarray}
\frac{d}{dt}\left(\frac{\partial L}{\partial \vec{v}_a}\right) - \frac{\partial L}{\partial \vec{r}_a} = 0.
\end{eqnarray}
The first term is rewritten as
\begin{eqnarray}
\frac{d}{dt}\left(\frac{\partial L}{\partial \vec{v}_a}\right) = m_a \dot{\vec{v}}_a,
\end{eqnarray}
and the second term becomes
\begin{eqnarray}
\frac{\partial L}{\partial \vec{r}_a} = -\sum_b\frac{P_bm_b}{\rho_b^2}\frac{\partial \rho_b}{\partial \vec{r}_a}+\sum_b\lambda_b\frac{h_b}{Dn_b}\frac{\partial n_b}{\partial \vec{r}_a}.
\end{eqnarray}
By differentiating $\rho_b$ and $n_b$ with respect to $\vec{r}_a$ and using Eq.(7) and Eq.(8), we have  
\begin{eqnarray}
\frac{\partial \rho_b}{\partial \vec{r}_a} &=& \sum_c m_c \vec{\nabla}_aW_{bc}(h_b)(\delta_{ba} - \delta_{ca}),\\
\frac{\partial n_b}{\partial \vec{r}_a} &=& \sum_c \vec{\nabla}_aW_{bc}(h_b)(\delta_{ba} - \delta_{ca}),
\end{eqnarray}
where $\delta_{ba}$ and $\delta_{ca}$ are the Kronecker delta. 
Using $\vec{\nabla}_aW_{ab}(h_a) = -\vec{\nabla}_bW_{ab}(h_a)$ and $W_{ab}(h_a) = W_{ba}(h_a)$, from Eqs.(18) through (20), we have
\begin{eqnarray}
\frac{\partial L}{\partial \vec{r}_a} = -\sum_b\frac{P_am_a}{\rho_a^2}\left(m_b-\Omega_a\right) \vec{\nabla}_aW_{ab}(h_a)-\sum_b\frac{P_bm_b}{\rho_b^2}\left(m_a-\Omega_b\right) \vec{\nabla}_aW_{ab}(h_b).
\end{eqnarray}
Here, $\Omega_b$ is the correction term derived from the variation of smoothing length as 
\begin{eqnarray}
\Omega_a \equiv \frac{\partial{\rho_a}}{\partial{h_a}}\left(\frac{Dn_a}{h_a}+\frac{\partial n_a}{\partial h_a}\right)^{-1}.
\end{eqnarray}
From Eq.(16), Eq.(17) and Eq.(21), we obtain the equation of motion of the following form,
\begin{eqnarray}
m_a \dot{\vec{v}}_a = -\sum_b\frac{P_am_a}{\rho_a^2}\left(m_b-\Omega_a\right) \vec{\nabla}_aW_{ab}(h_a)-\sum_b\frac{P_bm_b}{\rho_b^2}\left(m_a-\Omega_b\right) \vec{\nabla}_aW_{ab}(h_b).
\end{eqnarray}

%%%%%%%%%%%%%%%%%%%%%%%%%%%%%%%%%%

%%%%%%%%%%%%%%%%%%%%%%%%%%%%%%%%%%%%%%%%%%%%%%%%%%%%%%%%%%%%%%%%%%%%%%%%%%%%%%%%%%%%%%%%%%%%

%エネルギー方程式導出開始%%%%%%%%%%%%%%%%%%%%%%%%%%%%%%%%%%%%%%%%%%%%%%%%%%%%%%%%%%%%%%%%%%%%%
The equation of energy is given by two ways. We can derive it 
from the law of conservation of energy or the first law of thermodynamics. 
First we present the former. 

%エネルギー保存則から導く%%%%
The change of the internal energy is the same as that of the kinetic energy with an opposite sign. 
Thus the law of conservation is given by
\begin{eqnarray}
\left(\frac{dm_au_a}{dt}\right)_b+\left(\frac{dm_bu_b}{dt}\right)_a
 + \left(\frac{d}{dt}\frac{m_a|\vec{v}_a|^2}{2}\right)_b+\left(\frac{d}{dt}\frac{m_b|\vec{v}_b|^2}{2}\right)_a = 0.
\end{eqnarray}
From Eq.(23), the change of the kinetic energy due to the pairwise interaction between particles $a$ and $b$ is given by
\begin{eqnarray}
\left(\frac{d}{dt}\frac{m_a|\vec{v}_a|^2}{2}\right)_b&+&\left(\frac{d}{dt}\frac{m_b|\vec{v}_b|^2}{2}\right)_a =\vec{v}_a\cdot\frac{d}{dt}(m_a\vec{v}_a)+\vec{v}_b\cdot\frac{d}{dt}(m_b\vec{v}_b),
\nonumber\\
&=&-\vec{v}_a\cdot\left(\frac{P_am_a}{\rho_a^2}\left(m_b-\Omega_a\right) \vec{\nabla}_aW_{ab}(h_a)+\frac{P_bm_b}{\rho_b^2}\left(m_a-\Omega_b\right) \vec{\nabla}_aW_{ba}(h_b)\right)\nonumber\\
&&-\vec{v}_b\cdot\left(\frac{P_bm_b}{\rho_b^2}\left(m_a-\Omega_b\right) \vec{\nabla}_bW_{ba}(h_b)+\frac{P_am_a}{\rho_a^2}\left(m_b-\Omega_a\right) \vec{\nabla}_bW_{ab}(h_a)\right),\nonumber\\
&=&-\vec{v}_{ab}\cdot\frac{P_am_a}{\rho_a^2}\left(m_b-\Omega_a\right) \vec{\nabla}_aW_{ab}(h_a)\nonumber\\
&&-\vec{v}_{ab}\cdot\frac{P_bm_b}{\rho_b^2}\left(m_a-\Omega_b\right) \vec{\nabla}_aW_{ba}(h_a),\nonumber\\
\end{eqnarray}
where $\vec{v}_{ab} \equiv \vec{v}_a-\vec{v}_b$.
From Eq.(24) and Eq.(25) we obtain
\begin{eqnarray}
\left(\frac{dm_au_a}{dt}\right)_b+\left(\frac{dm_bu_b}{dt}\right)_a &=&  \vec{v}_{ab}\cdot\frac{P_am_a}{\rho_a^2}\left(m_b-\Omega_a\right) \vec{\nabla}_aW_{ab}(h_a)\nonumber\\
&&+\vec{v}_{ab}\cdot\frac{P_bm_b}{\rho_b^2}\left(m_a-\Omega_b\right) \vec{\nabla}_aW_{ab}(h_b).
\end{eqnarray}
By using constants $\alpha$ and $\beta$, this equation can be expressed in the following form
\begin{eqnarray}
\left(\frac{dm_au_a}{dt}\right)_b = \vec{v}_{ab}\cdot\alpha \frac{P_am_a}{\rho_a^2}\left(m_b-\Omega_a\right) \vec{\nabla}_aW_{ab}(h_a)+\vec{v}_{ab}\cdot\beta\frac{P_bm_b}{\rho_b^2}\left(m_a-\Omega_b\right) \vec{\nabla}_aW_{ab}(h_b),\nonumber\\
\end{eqnarray}
\begin{eqnarray}
\left(\frac{dm_bu_b}{dt}\right)_a &=& \vec{v}_{ab}\cdot(1-\alpha)\frac{P_am_a}{\rho_a^2}\left(m_b-\Omega_a\right) \vec{\nabla}_aW_{ab}(h_a)\nonumber\\
&&+\vec{v}_{ab}\cdot(1-\beta)\frac{P_bm_b}{\rho_b^2}\left(m_a-\Omega_b\right) \vec{\nabla}_aW_{ab}(h_b).
\end{eqnarray}
Then we replace $a$ with $b$ in Eq.(28).
\begin{eqnarray}
\left(\frac{dm_au_a}{dt}\right)_b& = &\vec{v}_{ab}\cdot(1-\alpha)\frac{P_bm_b}{\rho_b^2}\left(m_a-\Omega_b\right) \vec{\nabla}_bW_{ab}(h_b)\nonumber\\
&&+\vec{v}_{ab}\cdot(1-\beta)\frac{P_am_a}{\rho_a^2}\left(m_b-\Omega_a\right) \vec{\nabla}_bW_{ab}(h_a).
\end{eqnarray}
Eq.(29) must be the same as Eq.(27). 
Therefore we have the constraint $\alpha + \beta = 1$. 
In this paper we chose $(\alpha,\beta) = (1,0)$.
Thus the energy equation is given by
\begin{eqnarray}
m_a\frac{du_a}{dt} = \sum_b\frac{P_am_a}{\rho_a^2}\left(m_b-\Omega_a\right) \vec{v}_{ab}\cdot\vec{\nabla}_aW_{ab}.
\end{eqnarray}

Next we derive the equation of energy from the first law of thermodynamics. 
The law is given by 
\begin{eqnarray}
d(m_au_a) = -P_adV_a = \frac{m_aP_a}{\rho_a^2}d\rho_a.
\end{eqnarray}
We assume an isentropic dynamics. The equation of energy is given by
\begin{eqnarray}
m_a\frac{du_a}{dt} = -\frac{m_aP_a}{\rho_a^2}\frac{d\rho_a}{dt}.
\end{eqnarray}
By differentiating Eq.(8) with respect to $t$, we obtain
\begin{eqnarray}
\frac{d\rho_a}{dt} &=& \sum_b m_b \vec{v}_{ab}\cdot \vec{\nabla}_aW_{ab}(h_a) 
- \sum_b {m_b} \frac{h_a}{Dn_a}\partial_{h_a} W_{ab}(h_a)\frac{dn_a}{dt}\nonumber\\
&=&\sum_b m_b \vec{v}_{ab}\cdot \vec{\nabla}_{a}W_{ab}(h_a) 
- \frac{h_a}{Dn_a}\frac{\partial \rho_a}{\partial h_a}\frac{dn_a}{dt}.
\end{eqnarray}
We differentiate partially $n_a$ with respect to $t$ using Eq.(7) to derive the second term as
\begin{eqnarray}
\frac{dn_a}{dt} = \sum_b \vec{v}_{ab}\cdot \vec{\nabla}_aW_{ab}(h_a) 
-\frac{h_a}{Dn_a}\frac{\partial n_a}{\partial h_a}\frac{dn_a}{dt}.
\end{eqnarray}
By rewriting Eq.(34), the temporal differentiation of $n_a$ is given by
\begin{eqnarray}
\frac{dn_a}{dt} = \left(\sum_b \vec{v}_{ab}\cdot \vec{\nabla}_{a}W_{ab}(h_a)\right)\left(1+\frac{h_a}{Dn_a}\frac{\partial n_a}{\partial h_a}\right)^{-1}.
\end{eqnarray}
From Eqs.(22),(34) and (35) we obtain, 
\begin{eqnarray}
\frac{d\rho_a}{dt} = \sum_b (m_b-\Omega_a) \vec{v}_{ab}\cdot \vec{\nabla}_aW_{ab}(h_a).
\end{eqnarray}
When we substitute Eq.(36) into Eq.(31), we obtain the equation of energy.
\begin{eqnarray}
m_a\frac{du_a}{dt} = \sum_b \frac{P_am_a}{\rho_a^2}(m_b-\Omega_a) \vec{v}_{ab}\cdot \vec{\nabla}_aW_{ab}(h_a).
\end{eqnarray}
Eq.(37) is identical to Eq.(30). Thus it satisfies the law of conservation of energy.

%%%%%%%%%%%%%%%%%%%%%%%%%%%%%%%%%%%%%%%%%%%%%%%%%%%%%%%%%%%%%%%%%%%%%%%%%%%%%%%%%%%%%%%%%%%%

%EOS開始%%%%%%%%%%%%%%%%%%%%%%%%%%%%%%%%%%%%%%%%%%%%%%%%%%%%%%%%%%%%%%%%%%%%%%%%%%%%%%%%%%%%

We use the equation of state for an ideal gas,
\begin{eqnarray}
P_a = (\gamma -1) \rho_au_a.
\end{eqnarray}

%%%%%%%%%%%%%%%%%%%%%%%%%%%%%%%%%%%%%%%%%%%%%%%%%%%%%%%%%%%%%%%%%%%%%%%%%%%%%%%%%%%%%%%%%%%%

%SSPH定式化終了%%%%%%%%%%%%%%%%%%%%%%%%%%%%%%%%%%%%%%%%%%%%%%%%%%%%%%%%%%%%%%%%%%%%%%%%%%%%%

%問題と先行研究開始%%%%%%%%%%%%%%%%%%%%%%%%%%%%%%%%%%%%%%%%%%%%%%%%%%%%%%%%%%%%%%%%%%%%%%%%%%

\subsection{Problems of SSPH and Previous studies}
We can see that the density must be positive and continuous for the SSPH equations, 
Eqs.(8),(23) and (30) to be valid.
Thus, SSPH smooths density even at the density jump of a contact discontinuity.
As a result, the density in high (low) density region around the discontinuity is under (over) estimated. This error propagates to the evaluation of pressure through EOS. Eventually, we face a large pressure error around the contact discontinuity. This error works as an unphysical surface tension, resulting in the suppression of fluid instabilities (Agertz et al. 2007).
%%%先行研究%%%
To overcome this difficulty, several modifications of SSPH have been proposed.

%%%Price 2008%%%
Price (2008) proposed the use of AC. 
In his approach, AC makes the energy smooth so that the pressure distribution can be flat.
Thus both the density and the pressure become smooth.
It works fine if the discontinuity is due to the jump in the thermal energy.
However, it is not clear how we can handle the discontinuity in the chemical composition.

%%%DISPH%%%
Price (2008) and Read et al. (2010) showed that the KHI takes place with the smoothed pressure formulation which was developed by Ritchie \& Thomas (2001).  
Saitoh \& Makino (2013) clarified the mathematical and physical implication of the scheme used by Ritchie \& Thomas (2001) by means of a volume element $\Delta V_a$ given by $m_au_a/q_a$.
The quantity $q_a$ is the energy density defined by $q_a = \rho_au_a$, and 
it is proportional to the pressure in the case of the ideal gas. 
Thus, in DISPH, pressure should be positive and continuous. 
DISPH can deal with the contact discontinuity without any difficulty. 
However it still has some weak points. 
For example, it cannot evaluate the volume element of the fluid particle of 
which the pressure is very small (e.g. near the free surface of water).

%%%GSPH%%%
Cha et al. (2010) and Murante et al. (2011) adopted GSPH, which greatly improves
the pressure wiggles even in the shock tube test involving strong shock with the mach number of $10^5$.
GSPH was originally proposed by Inutsuka (2002).
They showed that the KHI grows well due to the good behavior of GSPH at contact discontinuities. It is, however, difficult to handle a non-ideal gas with GSPH, since one needs to solve a non-ideal Riemann problem and it is computationally expensive.

%%%高精度化%%%
Garc{\'{\i}}a-Senz et al. (2012) introduced an integral approach in order to evaluate the first derivative of the SPH approximation. This method can reduce the error of the gradient. With this method, they showed that the KHI grows even though they used the SSPH. Recently, Rosswog (2014) combined this method with DISPH. Although this method is very efficient, it is unclear how to deal with problems involving free surface.

%%%Ott%%%
Ott \& Schnetter (2003) proposed yet another method. 
In their method the volume element $\Delta V_a$ is defined as $1/n_a$, 
where $n_a$ is the number of density of particles. It is defined as
\begin{eqnarray}
n_a \equiv \sum_bW_{ab}(h_a).
\end{eqnarray}
This method performs well when the density evolution is continuous. 
If this is not the case, for example if two stars merge, 
this method would perform poorly, because the distribution of $n_a$ contains discontinuity.

We extend the idea of DISPH so that we can handle the region where pressure is quite low.
Here, we introduce a new quantity which is a virtual quantity and is not a physical one in order to evaluate a volume element.
In the following, we describe this new formulation of SPH.

%%%%%%%%%%%%%%%%%%%%%%%%%%%%%%%%%%%%%%%%%%%%%%%%%%%%%%%%%%%%%%%%%%%%%%%%%%%%%%%%%%%%%%%%%%%%

%新たなSPH%%%%%%%%%%%%%%%%%%%%%%%%%%%%%%%%%%%%%%%%%%%%%%%%%%%%%%%%%%%%%%%%%%%%%%%%%%%%%%%%%%

\section{New SPH}
In this section, we present the formulation of our new method, SPSPH. 
In SPSPH, we use the pseudo density $y$ and its associated pseudo mass $Z$, 
to obtain the volume element $\Delta V$. 
In addition, we let $y$ diffuse, following the diffusion equation to guarantee 
that its distribution is (or will be) smooth everywhere.
%一般拡散方程式定式化%%%%%%%%%%%%%%%%%%%%%%%%%%%%%%%%%%%%%%%%%%%%%%%%%%%%%%%%%%%%%%%%%%%%%%%%

\subsection{Equation for the pseudo density}
We derive the time evolution equation of the pseudo density $y$ with diffusion. 
First we drive the diffusion equation in the Eulerian view. 
We define the quantity $\vec{j}$ as the flux density of $Z$ by diffusion,
\begin{eqnarray}
\left(\frac{\partial y}{\partial t}\right)_{\rm{dif}} = -(\vec{\nabla} \cdot \vec{j})_{\rm{dif}}.
\end{eqnarray}

We derive $\vec{j}$ which satisfies the following property.
The quantity $\vec{j}$ depends on $\nabla y$, 
because it should reduce the jump in the distribution of $y$.
The simplest form is
\begin{eqnarray}
\vec{j} = -D_{\rm dif} \vec{\nabla} y.
\end{eqnarray}
The coefficient $D_{\rm dif}$ should be positive in Eq.(41). 
The diffusion equation of $y$ is
\begin{eqnarray}
\left(\frac{\partial y}{\partial t}\right)_{\rm{dif}} = D_{\rm dif}{\vec{\nabla}}^2 y + \vec{\nabla} D_{\rm dif}\cdot\vec{\nabla}{y}.
\end{eqnarray}
In this paper we use $D_{\rm dif}$ which is spatially constant. 
Thus the second term vanishes. 
In section 3.6, we discuss $D_{\rm dif}$ 
which depends on $\vec{r}$ or $y$.

Since $y$ should satisfy the continuity equation, 
the time evolution equation should contain the advection term as
\begin{eqnarray}
\frac{\partial y}{\partial t} = D_{\rm dif}{\vec{\nabla}}^2 y - \vec{\nabla}\cdot(y\vec{v}).
\end{eqnarray}

We then derive the equation for $y$ in the Lagrangian view. The Lagrangian derivative $d/dt$ is given by $d/dt = \partial/\partial t + \vec{v}\cdot\vec{\nabla}$. Therefore the equation is given by
\begin{eqnarray}
\frac{dy}{dt} = D_{\rm dif}{\vec{\nabla}}^2 y - y\vec{\nabla}\cdot{\vec{v}}.
\end{eqnarray}
The first term in the right-hand side of this equation indicates the evolution through diffusion and the second term indicates the change of pseudo density through compression or expansion. 
%%%%%%%%%%%%%%%%%%%%%%%%%%%%%%%%%%%%%%%%%%%%%%%%%%%%%%%%%%%%%%%%%%%%%%%%%%%%%%%%%%%%%%%%%%%%

%
%\newpage
%

%一般的定式化%%%%%%%%%%%%%%%%%%%%%%%%%%%%%%%%%%%%%%%%%%%%%%%%%%%%%%%%%%%%%%%%%%%%%%%%%%%%%%%%

\subsection{Generalized volume element and SPH equations}
In this section, we derive the set of fundamental equations of SPH based on the pseudo density $y$. 
We start with the volume element $\Delta V_a$ given by
\begin{eqnarray}
\Delta V_a = \frac{Z_a}{y_a},
\end{eqnarray}
where the pseudo density $y_a$ is given by
\begin{eqnarray}
y_a = \sum_b Z_b W_{ab}(h_a).
\end{eqnarray}
Here, the quantity $Z$ is an extensive quantity associated with $y$. 
In other words, $y$ is the spatial density of $Z$. We call $Z$ psuedo mass and $y$ pseudo density.

%運動方程式定式化開始%%%%%%%%%%%%%%%%%%%%%%%%%%%%%%%%%%%%%%%%%%%%%%%%%%%%%%%%%%%%%%%%%%%%%%%%

%ラグランジアン準備%%%%%%%%%%%%%%%

First we derive the equation of motion from the SPH Lagrangian. The Lagrangian is given by
\begin{eqnarray}
L(\vec{q}) = \sum_b \frac{1}{2}(m_b \vec{v}_b^2)-\sum_bm_bu_b + \sum_b \lambda_b\phi_b,
\end{eqnarray}
where
\begin{eqnarray}
\vec{q} &=& (\vec{r}_1,\vec{r}_2,\cdots,\vec{r}_b\cdots,h_1,h_2,\cdots,h_b,\cdots),\\
u_b &=& \frac{A_b}{\gamma-1} \left(\frac{m_by_b}{Z_b}\right)^{\gamma-1}.
\end{eqnarray}
The function $\phi$ is the constraint for the smoothing length 
and is the same as Eq.(12).

%%%%%%%%%%%%%%%%%%%%%%%%%%%%%%%%%%

%h空間での方程式変形%%%%%%%%%%%%%%%

As in section 2.1, by solving the Euler-Lagrange equation for $h_a$ and $\vec{r}_a$, 
we obtain the equation of motion in the following form
\begin{eqnarray}
m_a \dot{\vec{v}}_a = -\sum_b\frac{P_aZ_a}{y_a^2}\left(Z_b-\Omega_a\right) \vec{\nabla}_aW_{ab}(h_a)-\sum_b\frac{P_bZ_b}{y_b^2}\left(Z_a-\Omega_b\right) \vec{\nabla}_aW_{ba}(h_b).
\end{eqnarray}
Here, $\Omega_b$ is the correction term derived from the variation of smoothing length as 
\begin{eqnarray}
\Omega_a \equiv \frac{\partial{y_a}}{\partial{h_a}}\left(\frac{Dn_a}{h_a}+\frac{\partial n_a}{\partial h_a}\right)^{-1},
\end{eqnarray}
where
\begin{eqnarray}
\frac{\partial y_a}{\partial h_a}&=&\sum_bZ_b\partial_{h_a}W_{ab}(h_a).
\end{eqnarray}

We obtain the equation of energy from the equation of motion again as in section 2.1. 
It is given by 
\begin{eqnarray}
\frac{du_a}{dt} = \sum_b\frac{P_aZ_a}{y_a^2}\left(Z_b-\Omega_a\right) \vec{v}_{ab}\cdot\vec{\nabla}_aW_{ab} (h_a).
\end{eqnarray}

%%%%%%%%%%%%%%%%%%%%%%%%%%%%%%%%%%%%%%%%%%%%%%%%%%%%%%%%%%%%%%%%%%%%%%%%%%%%%%%%%%%%%%%%%%%%

%EOS開始%%%%%%%%%%%%%%%%%%%%%%%%%%%%%%%%%%%%%%%%%%%%%%%%%%%%%%%%%%%%%%%%%%%%%%%%%%%%%%%%%%%%

In the case of an ideal gas, pressure $P_a$ is given by
\begin{eqnarray}
P_a = (\gamma -1) \frac{m_au_ay_a}{Z_a}.
\end{eqnarray}

%%%%%%%%%%%%%%%%%%%%%%%%%%%%%%%%%%%%%%%%%%%%%%%%%%%%%%%%%%%%%%%%%%%%%%%%%%%%%%%%%%%%%%%%%%%%

%保存則開始%%%%%%%%%%%%%%%%%%%%%%%%%%%%%%%%%%%%%%%%%%%%%%%%%%%%%%%%%%%%%%%%%%%%%%%%%%%%%%%%%%
This formulation satisfies the law of conservation of energy. 
In addition, it conserves the linear momentum, the angular momentum and the total mass.
The temporal differentiation of the total linear momentum is given by
\begin{eqnarray}
\frac{d}{dt}\sum_am_a\vec{v}_a&=&\sum_a m_a\frac{d\vec{v}_a}{dt},\nonumber\\
&=&-\sum_a\sum_b\left[\frac{P_aZ_a}{y_a^2}\left(Z_b-\Omega_a\right) \vec{\nabla}_aW_{ab}(h_a)+\frac{P_bZ_b}{y_b^2}\left(Z_a-\Omega_b\right) \vec{\nabla}_aW_{ab}(h_b)\right].\nonumber\\
\end{eqnarray}
The sign of the right side changes, while the absolute value does not change, when we exchange the  indices $a$ and $b$, 
because the indices $a$ and $b$ are anticommutative in $\vec{\nabla}_{a}W_{ab}$. 
Thus the value of the right side is zero and it shows that our formulation conserves the  linear momentum. 
The temporal differentiation of the angular momentum is given by
\begin{eqnarray}
\frac{d}{dt}\sum_a\vec{r}_a\times m_a\vec{v}_a&=&\sum_a \vec{r}_a\times m_a\frac{d\vec{v}_a}{dt},\nonumber\\
&=&-\sum_a\sum_b\left[\frac{P_aZ_a}{y_a^2}\left(Z_b-\Omega_a\right) \vec{\tilde{r}}_{ab}\tilde{F}_{ab}(h_a)+\frac{P_bZ_b}{y_b^2}\left(Z_a-\Omega_b\right) \vec{\tilde{r}}_{ab}\tilde{F}_{ab}(h_b)\right],\nonumber\\
\end{eqnarray}
where the function $\tilde{F}_{ab}$ satisfies $\vec{r}_{ab}\tilde{F}_{ab}=\vec{\nabla}_aW_{ab}$ and $\tilde{\vec{r}}_{ab}$ equals $\vec{r}_a\times\vec{r}_b$. 
The  indices $a$ and $b$ are commutative in $\tilde{F}_{ab}$ and anticommutative in $\tilde{\vec{r}}_{ab}$. 
Thus the value of the right side is zero and it shows that our formulation conserves the angular momentum. 
In SPSPH we express the density of the particle $a$ as $\rho_a = m_ay_a/Z_a$. 
The total mass is given by 
\begin{eqnarray}
\sum_b \rho_b\Delta V_b &=& \sum_b \frac{m_by_b}{Z_b}\frac{Z_b}{y_b},\nonumber\\
&=&\sum_b m_b.
\end{eqnarray}
Therefore, SPSPH conserves the total mass because particle mass is the constant with respect to time.
%%%%%%%%%%%%%%%%%%%%%%%%%%%%%%%%%%%%%%%%%%%%%%%%%%%%%%%%%%%%%%%%%%%%%%%%%%%%%%%%%%%%%%%%%%%%

%一般的定式化終了

%%%%%%%%%%%%%%%%%%%%%%%%%%%%%%%%%%%%%%%%%%%%%%%%%%%%%%%%%%%%%%%%%%%%%%%%%%%%%%%%%%%%%%%%%%%%
 
%拡散方程式SPHバージョン%%%%%%%%%%%%%%%%%%%%%%%%%%%%%%%%%%%%%%%%%%%%%%%%%%%%%%%%%%%%%%%%%%%%%%

\subsection{Implementation of the diffusion term}

%yの拡散方程式導出%%%%%%%%%%%%%%%%%%%%%%%%%%%%%%%%%%%%%%%%%%%%%%%%%%%%%%%%%%%%%%%%%%%%%%%%%%%%
First, we derive the SPH expression of the first term in the right side of Eq.(44) for SPH.
The SPH expression for the Laplacian has the following form (e.g. Brookshaw 1985):
\begin{eqnarray}
{\vec{\nabla}}^2 A_a = 2\sum_b (A_a-A_b) \Delta V_b \frac{\vec{r}_{ab}\cdot \vec{\nabla}_aW_{ab}(h_a)}{|\vec{r}_{ab}|^2}.
\end{eqnarray}
 Therefore, the diffusion term of $y$ is given by
\begin{eqnarray}
\left(\frac{dy_a}{dt}\right)_{\rm{dif}} 
=2D_{\rm{dif}}\sum_b (y_a-y_b) \frac{Z_b}{y_b} \frac{\vec{r}_{ab}\cdot \vec{\nabla}_aW_{ab}(h_a)}{|\vec{r}_{ab}|^2}.
\end{eqnarray}
%Zの拡散方程式導出%%%%%%%%%%%%%%%%%%%%%%%%%%%%%%%%%%%%%%%%%%%%%%%%%%%%%%%%%%%%%%%%%%%%%%%%%%%%

What we actually need is the equation for the pseudo mass $Z$, not for $y$, 
since we obtain $y$ from $Z$ using Eq.(46). 
In the following, we derive the diffusion equation of $y$ for $Z$. 
The requirement is that diffusion doesn't change the Lagrangian.
Therefore $Z$ must evolve so that the volume of particle is not changed by the diffusion of $y$, because Lagrangian $L(y,Z)$ has the form $L(Z/y)$. Therefore, 
\begin{eqnarray}
\left(\frac{d\frac{Z}{y}}{dt}\right)_{\rm dif}=\frac{1}{y}\left(\frac{dZ}{dt}\right)_{\rm dif}-\frac{Z}{y^2}\left(\frac{dy}{dt}\right)_{\rm dif}=0.
\end{eqnarray}
Thus the diffusion equation for $Z$ is expressed as
\begin{eqnarray}
\left(\frac{dZ_a}{dt}\right)_{\rm{dif}} 
=\frac{Z_a}{y_a}\left(\frac{dy_a}{dt}\right)_{\rm{dif}}
=2D_{\rm{dif}}\frac{Z_a}{y_a}\sum_b (y_a-y_b) \frac{Z_b}{y_b} \frac{\vec{r}_{ab}\cdot \vec{\nabla}_aW_{ab}(h_a)}{|\vec{r}_{ab}|^2},
\end{eqnarray}
and it gives to the equation of time evolution of $Z$.
\begin{eqnarray}
\frac{dZ_a}{dt}
=2D_{\rm{dif}}\frac{Z_a}{y_a}\sum_b (y_a-y_b) \frac{Z_b}{y_b} \frac{\vec{r}_{ab}\cdot \vec{\nabla}_aW_{ab}(h_a)}{|\vec{r}_{ab}|^2}.
\end{eqnarray}

%%%%%%%%%%%%%%%%%%%%%%%%%%%%%%%%%%%%%%%%%%%%%%%%%%%%%%%%%%%%%%%%%%%%%%%%%%%%%%%%%%%%%%%%%%%%

%タイムステップ導出%%%%%%%%%%%%%%%%%%%%%%%%%%%%%%%%%%%%%%%%%%%%%%%%%%%%%%%%%%%%%%%%%%%%%%%%%%

\subsection{Time step criterion and stability analysis}
　\ The time step for integration is limited by the Courant condition for numerical stability. Since we use the leap-frog method for time integration, the time step $\Delta t_{\rm CFL}$ is given by
%maximum：とりうる中で最大のということ。
\begin{eqnarray}
\Delta t_{\rm CFL} &=& \min_{a} dt_a,\\
dt_a &=& C_{\rm CFL}\frac{2h_a}{\max_bv_{ab}^{\rm sig}}.
\end{eqnarray}
It is used as the time step for SPH without diffusion. 
In this paper, we set $C_{\rm CFL} = 0.6$. 
We need to derive the maximum time step $\Delta t_{\rm dif}$ for the diffusion equation of SPH, 
because it can become smaller than $\Delta t_{\rm CFL}$.

%線型安定性解析開始%%%%%%%%%%%%%%%%%%%%%%%%%%%%%%%%%%%%%%%%%%%%%%%%%%%%%%%%%%%%%%%%%%%%%%%%%%
We derive the time step for the diffusion equation of SPH by the linear stability analysis of Eq.(61). 
We perturb the pseudo density $y$ and the pseudo mass $Z$ from the uniform state 
which satisfies $y_a=y^0, Z_a=Z^0$ and $h_a=h^0$ for all particles $a$. 
The perturbations of $y$ and $Z$ are defined as $\delta y$ and $\delta Z$. 
By perturbing Eq.(61), it is given by 
\begin{eqnarray}
\left[\frac{d(Z_a+\delta Z_a)}{dt}\right] = 2D_{\rm dif}\frac{Z_a+\delta Z_a}{y_a+\delta y_a}\sum_b(y_a-y_b+\delta y_a-\delta y_b)\frac{Z_b+\delta Z_b}{y_b + \delta y_b}\frac{\vec{r}_{ab}\cdot\vec{\nabla}_aW_{ab}(h_a)}{|\vec{r}_{ab}|^2}.\nonumber\\
\end{eqnarray}
We start analysis by $y_a=y^0, Z_a=Z^0$ and $h_a=h^0$ for all $a$. Thus
\begin{eqnarray}
\left[\frac{d(Z^0+\delta Z_a)}{dt}\right] = 2D_{\rm dif}\frac{Z^0+\delta Z_a}{y^0+\delta y_a}\sum_b(y^0-y^0+\delta y_a-\delta y_b)\frac{Z^0+\delta Z_b}{y^0 + \delta y_b}\frac{\vec{r}_{ab}\cdot\vec{\nabla}_aW_{ab}(h^0)}{|\vec{r}_{ab}|^2}.\nonumber\\
\end{eqnarray}
The perturbation equation to the first order is expressed as
\begin{eqnarray}
\frac{d(\delta Z_a)}{dt}= 2D_{\rm dif}\frac{Z^0}{y^0}\sum_b(\delta y_a-\delta y_b)\frac{Z^0}{y^0}\frac{\vec{r}_{ab}\cdot\vec{\nabla}_aW_{ab}}{|\vec{r}_{ab}|^2}.\end{eqnarray}
We can express $W_{ab}(h^0)$ as $W_{ab}$ for convenience, because $h^0$ is the same for all particles. 
In the following, we consider the simplest case of a 1D problem 
in which particles are placed in equal spacing. 
In this case, $\delta Z_a$ can be expressed by the fourier series as follows
\begin{eqnarray}
\delta Z_a(r) = \sum_l A_l e^{\omega t}e^{ik_lr_a},
\end{eqnarray}
where $\omega$ is a complex number. Consider the perturbation of $Z$ with the wave number $k$. 
\begin{eqnarray}
\delta Z_{a,k}(r) = Ae^{\omega t}e^{ikr_a}.
\end{eqnarray}
The perturbation of $y$ is then given by
\begin{eqnarray}
\delta y_{a,k}(r) = \sum_c A_me^{\omega t}e^{ikr_c}W_{ac},
\end{eqnarray}
and Eq.(67) with the wave number $k$ becomes
\begin{eqnarray}
\frac{(e^{\omega(t+\Delta t_{\rm dif})}-e^{\omega t})e^{ikr_a}}{\Delta t_{\rm dif}}=
2D_{\rm dif}\frac{Z^0}{y^0}\sum_b\left(\sum_c e^{\omega t}e^{ikr_c}W_{ac}-\sum_c e^{\omega t}e^{ikr_c}W_{bc}\right)\frac{Z^0}{y^0}\frac{\vec{r}_{ab}\cdot\vec{\nabla}W_{ab}}{r_{ab}^2}.\nonumber\\
\end{eqnarray}
We divide the equation by $\exp({\omega t})\exp({ikr_a})$ and use that the particle spacing equals $Z^0/y^0$. Thus, we obtain
\begin{eqnarray}
\frac{(e^{\omega\Delta t_{\rm dif}}-1)}{\Delta t_{\rm dif}}&=&
2D_{\rm dif}\frac{Z^0}{y^0}\sum_b\left(\sum_c e^{ikr_{ca}}W_{ac}-\sum_c e^{ik(r_{cb}-r_{ba})}W_{bc}\right)\frac{Z^0}{y^0}\frac{\vec{r}_{ab}\cdot\vec{\nabla}W_{ab}}{r_{ab}^2},\nonumber\\
&=&2D_{\rm dif}\frac{Z^0}{y^0}\sum_b\left[\fourier{W}-\fourier{W}e^{-ikr_{ba}}\right]\frac{\vec{r}_{ab}\cdot\vec{\nabla}W_{ab}}{r_{ab}^2},\nonumber\\
&=&2D_{\rm dif}\fourier{W}\frac{Z^0}{y^0}\sum_b\left(1-e^{-ikr_{ba}}\right)\frac{\vec{r}_{ab}\cdot\vec{\nabla}W_{ab}}{r_{ab}^2},
\end{eqnarray}
where $\mathcal{F}({f})$ expresses the fourier transform of a function $f$.
We used $W_{ac} = W_{ca}$.
Note that Eq.(72) converges to zero under $a=b$ in the real space in the limit of $r_{ab}\rightarrow 0$. 
The lowest order of the function $\vec{\nabla}_aW_{ab}$ respect to $r_{ab}$ is $\mathcal{O}(0)$. 
Thus, that of ${\vec{r}_{ab}\cdot\vec{\nabla}W_{ab}}/{r_{ab}^2}$ is $\mathcal{O}(r_{ab}^{-1})$. 
If $[{1-\exp({ikr_{ab}})}]$ converges to zero faster than $\mathcal{O}(r_{ab})$, 
Eq.(72) converges to zero. 
From l'Hopital's rule, we obtain
\begin{eqnarray}
{\rm Re}\left[\lim_{r_{ab}\rightarrow 0} \frac{1-\exp({ikr_{ab}})}{\mathcal{O}(r_{ab}^{1})}\right] = \lim_{r_{ab}\rightarrow 0} \frac{-\sin kr_{ab}}{\mathcal{O}(0)} = 0.
\end{eqnarray}
Thus, Eq.(72) converges to zero for $r_{ab}\rightarrow 0$ in the real space. 
Therefore, by defining the function $g$ as
\begin{eqnarray}
g(r_{ab}) \equiv \left\{
    \begin{array}{ll}
      \sum_c {\frac{\vec{r}_{ac}\cdot\vec{\nabla}W_{ac}}{r_{ac}^2}}& (b=a) \\
      - {\frac{\vec{r}_{ab}\cdot\vec{\nabla}W_{ab}}{r_{ab}^2}}& (b\neq a)
    \end{array}
  \right.,
\end{eqnarray}
we can obtain the following equation:
\begin{eqnarray}
\frac{e^{\omega\Delta t_{\rm dif}}-1}{\Delta t_{\rm dif}}= 2D_{\rm dif}\fourier{g}\fourier{W}.
\end{eqnarray}

The condition that the perturbation damps is a positive $\Delta t_{\rm dif}$ 
that satisfies $|\exp({\omega\Delta t_{\rm dif}})| < 1$ exists for all possible values of $k$.
Therefore the stability condition is 
\begin{eqnarray}
0 < {\Delta t_{\rm dif}} < \min_k\frac{-1}{D_{\rm dif}\fourier{g}\fourier{W}}.
\end{eqnarray}

%%%%%%%%%%%%%%%%%%%%%%%%%%%%%%%%%%%%%%%%%%%%%%%%%%%%%%%%%%%%%%%%%%%%%%%%%%%%%%%%%%%%%%%%%%%%

The upper limit of the wave number $k_{\rm max}$ is $2\pi/\Delta x$ where $\Delta x$ is the particle spacing.
From Eq.(76), the function $\fourier{g}\fourier{W}$ must be negative for all possible values of $k$. 
First, we show that $\fourier{g}$ is negative. 
The kernel function $W(|\vec{r}_{ab}|)$ takes the maximam value at $|\vec{r}_{ab}| = 0$, 
and is decreasing for both sides. 
Therefore, the gradiant is positive for $\vec{r}_{ab} < 0$, and negative for $\vec{r}_{ab} > 0$. 
All terms in the summation of $g(r_{ab})$ is negative, and therefore $g(r_{ab})$ is negative. 
$\fourier{g}$ is given by
\begin{eqnarray}
\fourier{g} &=& \frac{Z^0}{y^0}g(r_{ab}) e^{ikr_{ab}},\nonumber\\
&=&\sum_{b(a\neq b)}\frac{Z^0}{y^0}\left(1-e^{ikr_{ab}}\right)\frac{r_{ab}\cdot\vec{\nabla}W_{ab}}{r_{ab}^2}.
\end{eqnarray}
The quantity $1-\exp({ikr_{ab}})$ is positive, whereas $\fourier{g}$ is negative. 
Thus the function $\fourier{W}$ must be positive for the possible $k$. 
In this paper we use the Wendland $C^2$ and $C^4$ function for one dimension and two dimensions, 
whose fourier transform are positive for all wave number 
$0\leq k \leq k_{\rm max}$ as kernel function, as proposed by Dehnen \& Aly (2012).
We note that the perturbation does not grow within a finite time even though the fourier transform of these function can become non-positive for an inhomogeneous particle distribution, since $dZ/dt$ is small enough.

%%%%%%%%%%%%%%%%%%%%%%%%%%%%%%%%%%%%%%%%%%%%%%%%%%%%%%%%%%%%%%%%%%%%%%%%%%%%%%%%%%%%%%%%%%%%
%拡散方程式はステップ数を増やす%%%%%%%%%%%%%%%%%%%%%%%%%%%%%%%%%%%%%%%%%%%%%%%%%%%%%%%%%%%
The maximum value of $\Delta t_{\rm dif}$ can become very small. 
Therefore, we consider the option to use separate time steps for the diffusion equation and the hydrodynamics. 
We use $\Delta t_{\rm CFL}$ for the hydrodynamics and use $\Delta t_{\rm dif}^{\rm imp}$ for the diffusion in practice. The time step $\Delta t_{\rm dif}^{\rm imp}$ is given by
\begin{eqnarray}
\Delta t_{\rm dif}^{\rm imp} = \frac{\Delta t_{\rm CFL}}{M},
\end{eqnarray}
where,
\begin{eqnarray}
M = \left\lceil \frac{\Delta t_{\rm CFL}}{\Delta t_{\rm dif}} \right\rceil.
\end{eqnarray}
We use $\Delta t_{\rm dif}$ as  
\begin{eqnarray}
\Delta t_{\rm dif} = \frac{(\min_{ab} |\vec{r}_{ab}|)^2}{(2D_{\rm dif})},
\end{eqnarray} 
It is easy to show that this is the maximum time step for the equally spaced particles in the one dimension case. 
Dehnen \& Aly (2012) showed $\fourier {W} \leq 1$. 
From Eq.(74),
\begin{eqnarray}
\fourier{g} &<& 2\sum_b \frac{Z^0}{y^0}{\frac{\vec{r}_{ab}\cdot\vec{\nabla}W_{ab}}{r_{ab}^2}},\nonumber\\
&\simeq& 2\sum_b \frac{Z^0}{y^0}\frac{W_{ab}}{r_{ab}^2},\nonumber\\
&<& \frac{2}{\min_{b}r_{ab}^2}.
\end{eqnarray}
Thus the maximum value of the function $\fourier{g}$ respect to $r_{ab}$ satisfies 
\begin{eqnarray}
\max\fourier{g} < \frac{2}{\min_{ab}r_{ab}^2},
\end{eqnarray}
and we have Eq.(80).

Figure 1 shows that the timestep obtained using of Eq.(80) is smaller than that using of Eq.(76). 
Thus, the time step derived from Eq.(80) satisfies the condition of Eq.(76), independent of the value of the diffusion coefficient, because both equations depend on 
inverse proportion of diffusion coefficient. 
Therefore we use the time step derived from Eq.(80).
\begin{figure}[htbp]
\begin{center}
	\FigureFile(80mm,50mm){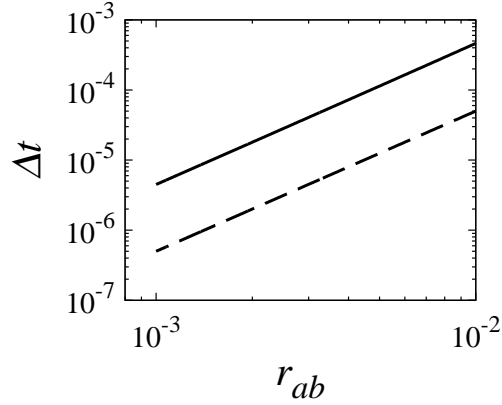}
\end{center}
\caption{Time steps derived from Eq.(76) and Eq.(80).
 The solid line shows time steps evaluated by Eq.(76) and the dashed line represents those evaluated by Eq.(80).}
\end{figure}
%%%%%%%%%%%%%%%%%%%%%%%%%%%%%%%%%%%%%%%%%%%%%%%%%%%%%%%%%%%%%%%%%%%%%%%%%%%%%%%%%%%%%%%%%%%%

%人工粘性開始%%%%%%%%%%%%%%%%%%%%%%%%%%%%%%%%%%%%%%%%%%%%%%%%%%%%%%%%%%%%%%%%%%%%%%%%%%%%%%%
\subsection {Artificial Viscosity}

To deal with shocks, we need to add an artificial viscosity term to the energy equation and the equation of motion. The viscosity term for the equation of motion is
\begin{eqnarray}
\left(\frac{dm_a\vec{v}_a}{dt}\right)_{\rm AV} = -\sum_b{m_a}{m_b}F_{ab}\Pi_{ab}\frac{\left[\vec{\nabla}W_{ab}(h_a)+\vec{\nabla}W_{ab}(h_b)\right]}{2},
\end{eqnarray}
and that for the energy equation is
\begin{eqnarray}
\left(\frac{dm_au_a}{dt}\right)_{\rm AV} = \frac{1}{2}\sum_b{m_a}{m_b}F_{ab}\Pi_{ab} \vec{v}_{ab}\cdot\frac{\left[\vec{\nabla}W_{ab}(h_a)+\vec{\nabla}W_{ab}(h_b)\right]}{2}.
\end{eqnarray}

Here, $\Pi_{ab}$ gives the viscosity and $F_{ab}$ is a ``switch'' function to reduce shear viscosity (Balsara 1995). 
There are several different forms of the function $\Pi_{ab}$. In this paper we adopt the function proposed by Monaghan (1997). It is expressed as
\begin{eqnarray}
\Pi_{ab} = -\alpha^{\rm vis}\frac{v_{ab}^{\rm sig}\omega_{ab}}{(\rho_a+\rho_b)}H(-\vec{v}_{ab}\cdot\vec{r}_{ab}),
\end{eqnarray}
where $H(x)$ is the Heaviside function, $v_{ab}^{\rm sig}=cs_{a}+cs_{b}-3\omega_{ab}$ , $\omega_{ab} = \vec{v}_{ab}\cdot\vec{r}_{ab}/\vec{r}_{ab}$ and $cs$ is the sound speed. 
In this paper we set $\alpha^{\rm vis} = 1$. 
The mass density $\rho$ is calculated by $\rho=m/\Delta V$. 
We used the standard Balsara switch (Balsara 1995) for $F_{ab}$. It is given by
\begin{eqnarray}
F_{ab} &=& \frac{1}{2}(F_a+F_b),\\
F_{a} &=& \frac{|\vec{\nabla}\cdot\vec{v}_{a}|}{|\vec{\nabla}\cdot\vec{v}_a|+|\vec{\nabla}\times\vec{v}_a|+\epsilon_bcs_a/h_a},
\end{eqnarray}
where $\epsilon_{b}$ is a small constant and in this paper we set $\epsilon_{b}=10^{-4}$ to prevent numerical overflow. 
The divergence and rotation of velocity are
\begin{eqnarray}
\vec{\nabla}\cdot\vec{v}_a &=& -\sum_b\frac{Z_b}{y_a}\vec{v}_{ab}\cdot\vec{\nabla}W_{ab}(h_a),\\
\vec{\nabla}\times\vec{v}_a &=& -\sum_b\frac{Z_b}{y_a}\vec{v}_{ab}\times\vec{\nabla}W_{ab}(h_a).
\end{eqnarray}
In SSPH, we used density $\rho$ and mass $m$ in place of $y$ and $Z$.
%%%%%%%%%%%%%%%%%%%%%%%%%%%%%%%%%%%%%%%%%%%%%%%%%%%%%%%%%%%%%%%%%%%%%%%%%%%%%%%%%%%%%%%%%%%%

\subsection{Possible choice of the diffusion coefficient}

So far we assume that the diffusion constant, $D_{\rm dif}$ is actually constant in space. 
However, it is possible to change $D_{\rm dif}$, since the diffusion itself does not introduce  numerical error. 
In the limit of $D_{\rm dif}\rightarrow 0$, our new SPH scheme is reduced back to SSPH. 
Thus, where SSPH works fine, or in other words, after $y$ has become reasonably smooth, 
it makes sense to reduce $D_{\rm dif}$ so that we can increase $\Delta t_{\rm dif}$. 
For example, consider a polynomial form expressed as 
\begin{eqnarray}
D_{\rm dif}=\sum_n B_n |\vec{\nabla} y|^{\mu_n},
\end{eqnarray}
where $B_n$ and $\mu_n$ are positive actual constants. Then Eq.(42) becomes
\begin{eqnarray}
\left(\frac{\partial y}{\partial t}\right)_{\rm{dif}} = 
\sum_n B_n|\vec{\nabla}y|^{\mu_n-1}\left(|\vec{\nabla}y|{\vec{\nabla}}^2 y + \mu_n \vec{\nabla}|\vec{\nabla} y|\cdot\vec{\nabla} y\right).
\end{eqnarray}

However, in this paper, we use the diffusion coefficient $D_{\rm dif}$ which is constant in space and variable in time. 
One simple choice of $D_{\rm dif}$ is such that $\Delta t_{\rm dif} = \Delta t_{\rm CFL}$. 
Such $D_{\rm dif}$ is given by $D_M$ which is defined as
\begin{eqnarray}
D_M \equiv \frac{(\Delta x)^2}{2\Delta t_{\rm CFL}},
\end{eqnarray}
where $\Delta x$ is the minimum particle spacing. In this paper, we use the initial particle separator as $\Delta x$, since the density change in our tests is small. If one uses this method for more realistic cases, it is necessary to adopt the minimum particles distribution at each time step and/or to evaluate the diffusion coefficient for each particle.
For example,
\begin{eqnarray}
D_{M,i} = \frac{\Lambda h_i^2}{2 \Delta t_{\rm CFL}},
\end{eqnarray}
where $\Lambda$ is a positive constant and equals to about $2.8\eta$ with the Wendland $C^4$ kernel for two dimentions.

%%%%%%%%%%%%%%%%%%%%%%%%%%%%%%%%%%%%%%%%%%%%%%%%%%%%%%%%%%%%%%%%%%%%%%%%%%%%%%%%%%%%%%%%%%%%

%色々テスト開始%%%%%%%%%%%%%%%%%%%%%%%%%%%%%%%%%%%%%%%%%%%%%%%%%%%%%%%%%%%%%%%%%%%%%%%%%%%%%%

\section{Numerical tests}

In this section, we compare the results of SPSPH to those of SSPH. 
In all tests, we found SPSPH gives better, or at least similar, results, 
compared to those of SSPH.

In section 4.1, we show the results of the ``Square test'' proposed by Saitoh \& Makino (2013), 
in which the evolution of a square-shaped high-density fluid embedded in a low-density fluid is solved. 
We then show the results of the shock tube tests in section 4.2
and those of the KHI tests in section 4.3.
We give the results of square tests with an extreme density contrast in section 4.4.
Finally, we show the results of the one-dimensional hydrostatic equilibrium tests and discuss the remaining difficulty of SPH induced by inhomogeneous particle distribution.
%ゆるめの静水圧平衡%%%%%%%%%%%%%%%%%%%%%%%%%%%%%%%%%%%%%%%%%%%%%%%%%%%%%%%%%%%%%%%%%%%%%%%%%%

\subsection{The square test}

　\ This test was first used by Saitoh \& Makino (2013). 
The computational domain is a two-dimensional square of unit size, 
$-0.5\leq x < 0.5$ and $-0.5\leq z < 0.5$, with the periodic boundary condition. 
The initial density is given by
\begin{eqnarray}
\left\{
\begin{array}{l}
\rho = 4 \ \ -0.25\leq x\leq 0.75\  {\rm and}\  0.25\leq z \leq 0.75,\\
\rho = 1 \ \ {\rm otherwise}.
\end{array}
\right.
\end{eqnarray}
We set $\gamma = 5/3$ and initially $P=2.5$, ${v}_x = {v}_z = 0$.

We used two different initial distributions for $y$. 
In the first one, the initial value of pseudo-density $y_{\rm init}$ is unity everywhere.
The second one is that the initial pseudo-density is the same as the initial density, $i.e., y_{\rm init} = \rho$. 
In this way, we can see if  
our scheme can handle initial discontinuity of $y$.

We expressed the density difference in two ways. 
In the first one, particles in the two regions have the same mass and 
the spacing between particles is changed. 
Thus the number density of particle is different in the two regions. 
The particle mass in this case is set to 1/16276.
In the second one, the particle mass is changed so that the particle spacing can be the same in the whole region.
The particle mass for the high density region is 1/4069 while that for the low density region is 1/16276.
We summarize our runs in table 1.

\begin{table}[htb]
  \begin{center}
    \caption{Summary of the runs for the square test.}
    \begin{tabular}{lcrrr}
      Model & SPH scheme & Particle mass & $y_{\rm init}$ & $D_{\rm dif}$ \\ \hline \hline
      run 1 & SSPH & equal & - & -\\
      run 2 & SPSPH & equal & 1 & $D_{M}$\\
      run 3 & SPSPH & equal & density & $10D_{M}$\\ 
      run 3a & SPSPH & equal & density & $D_{M}$\\ \hline
      run 4 & SSPH & unequal & - & -\\
      run 5 & SPSPH & unequal & 1 & $D_{M}$\\
      run 6 & SPSPH & unequal & density & $D_{M}$\\ \hline
    \end{tabular}
  \end{center}
\end{table}

For run 3 we let $y$ diffuse for 0.1 unit time with the diffusion coefficient of $500(\Delta x)^2$, before we start the calculation. 
Here, $\Delta x$ is the initial particle separation in the high density region. 
In addition, we set $D_{\rm dif} = 10D_{M}$ for run 3 as is seen in table 1. 
This is because the diffusion is insufficient in the case of $D_{\rm dif} = D_{M}$ (run 3a) and run 3a gives an inadequate result. One might consider that the use of 10 $D_M$ results in a high computationally cost. However, the actual computational cost of run 3 is only twice as expensive as that of run 3a. We consider that this increase of the computational cost is much smaller when we simulate complicate scientific phenomena in realistic situations.

%%%%結果%%%%
Figures 2 and 3 show the time evolution up to $t = 8$. 
It is clear that SPSPH can handle the contact discontinuity quite well 
while SSPH cannot. 
SSPH made the square region almost completely circular. 

Figures 4 and 5 show the cross section of the pressure and the acceleration for SSPH and SPSPH. 
The SSPH results show large errors at boundary. These errors in SSPH work as the unphysical surface tension (Saitoh \& Makino 2013).
As a result, SSPH cannot maintain the initial particle distribution (figures 2 and 3).

The results with SPSPH are far better than those with SSPH, 
as expected, even when $y$ is initially discontinuous. 
The reason for this behavior is that it requires the continuity of $y$ 
instead of that of the density. 
Even when $y$ is initially discontinuous, SPSPH can make it continuous and smooth. 
Figure 6 shows the cross section of the density (SSPH) 
and the pseudo density (SPSPH). 
We can see that pseudo density $y$ has become pretty smooth in SPSPH, 
while the jump in the density is well maintained.
It indicates that SPSPH performs well even when the initial distribution of $y$ contains discontinuity.

%密度比1:4の図開始

\begin{figure}[htbp]
\begin{center}
  	\FigureFile(160mm,50mm){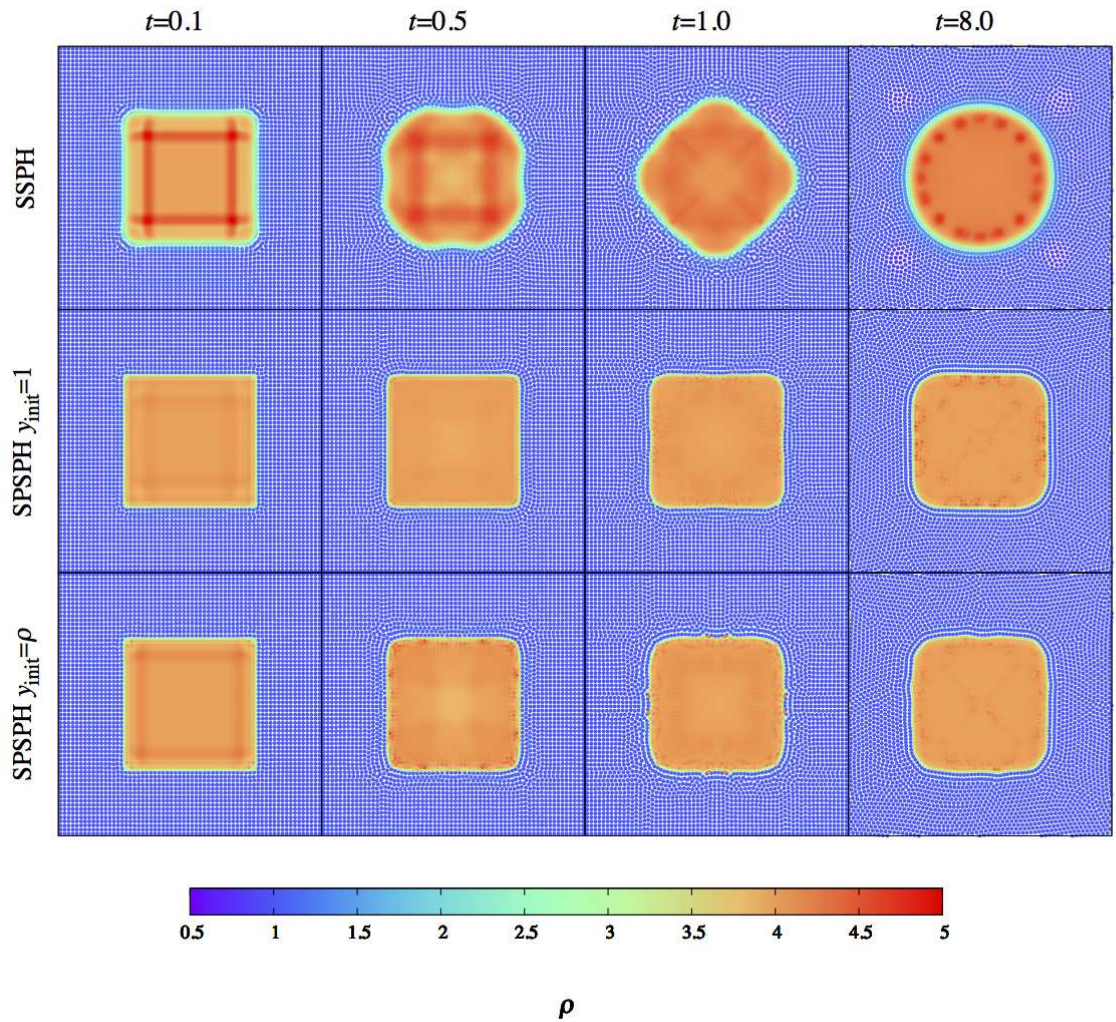}
\end{center}
\caption{Results of the square test, for runs 1 to 3 in table 1 from top to bottom, snapshot at $t=0.1, 0.5, 1.0, 8.0$ are shown. In this test the density ratio is 4:1 with the mass ratio of 1:1. 
The top row is the result of SSPH, 
while the middle and bottom rows are results of SPSPH in which $y_{\rm init}$ is 1 and density, respectively.}
\end{figure}
\begin{figure}[htbp]
\begin{center}
	\FigureFile(160mm,50mm){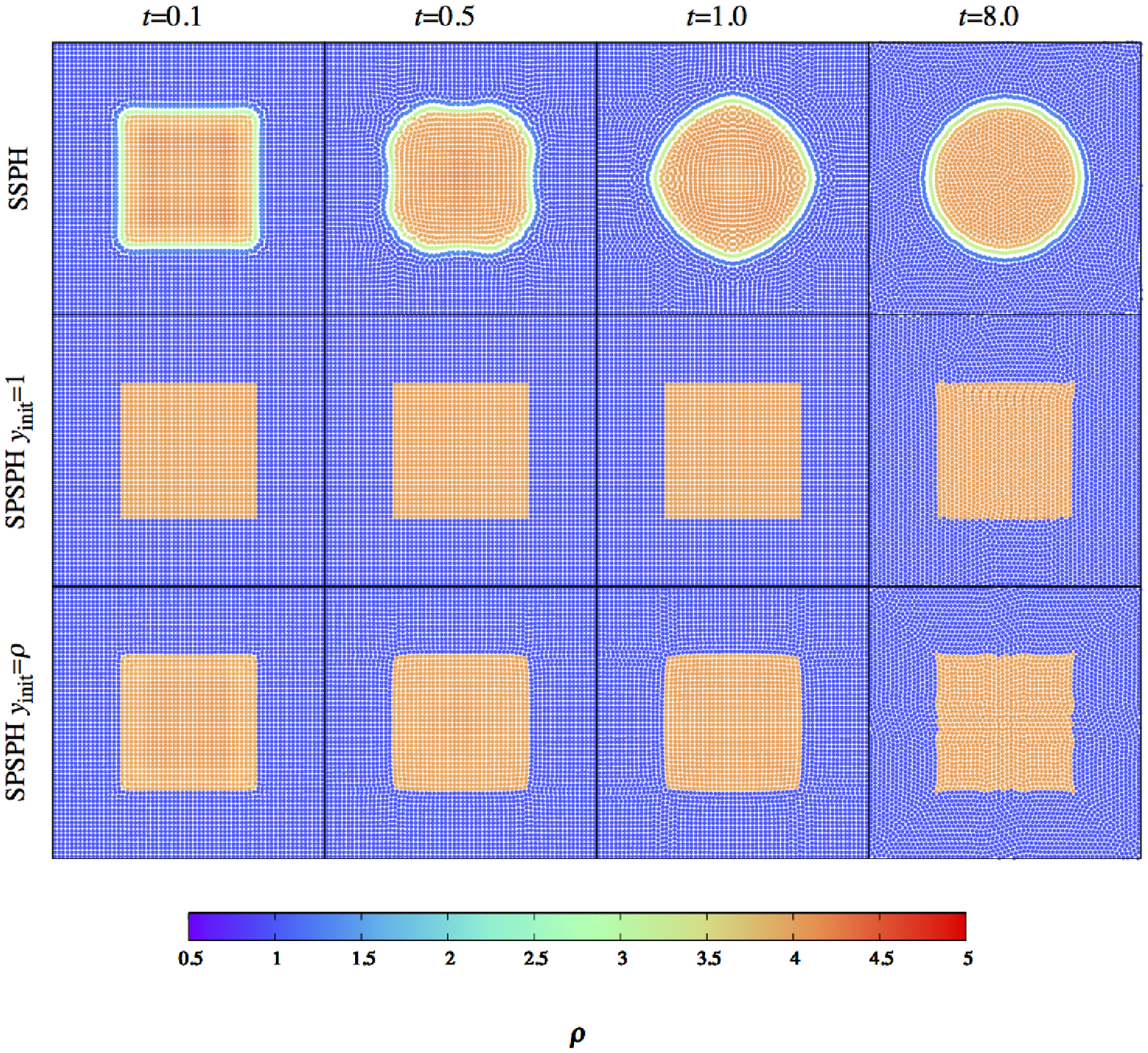}
\end{center}
\caption{The same as Fig. 2 but for runs 4 to 6 and with the mass ratio of 4:1.}
\end{figure}

%密度比1:4の図終了

\begin{figure}[htbp]
\begin{center}
\FigureFile(160mm,50mm){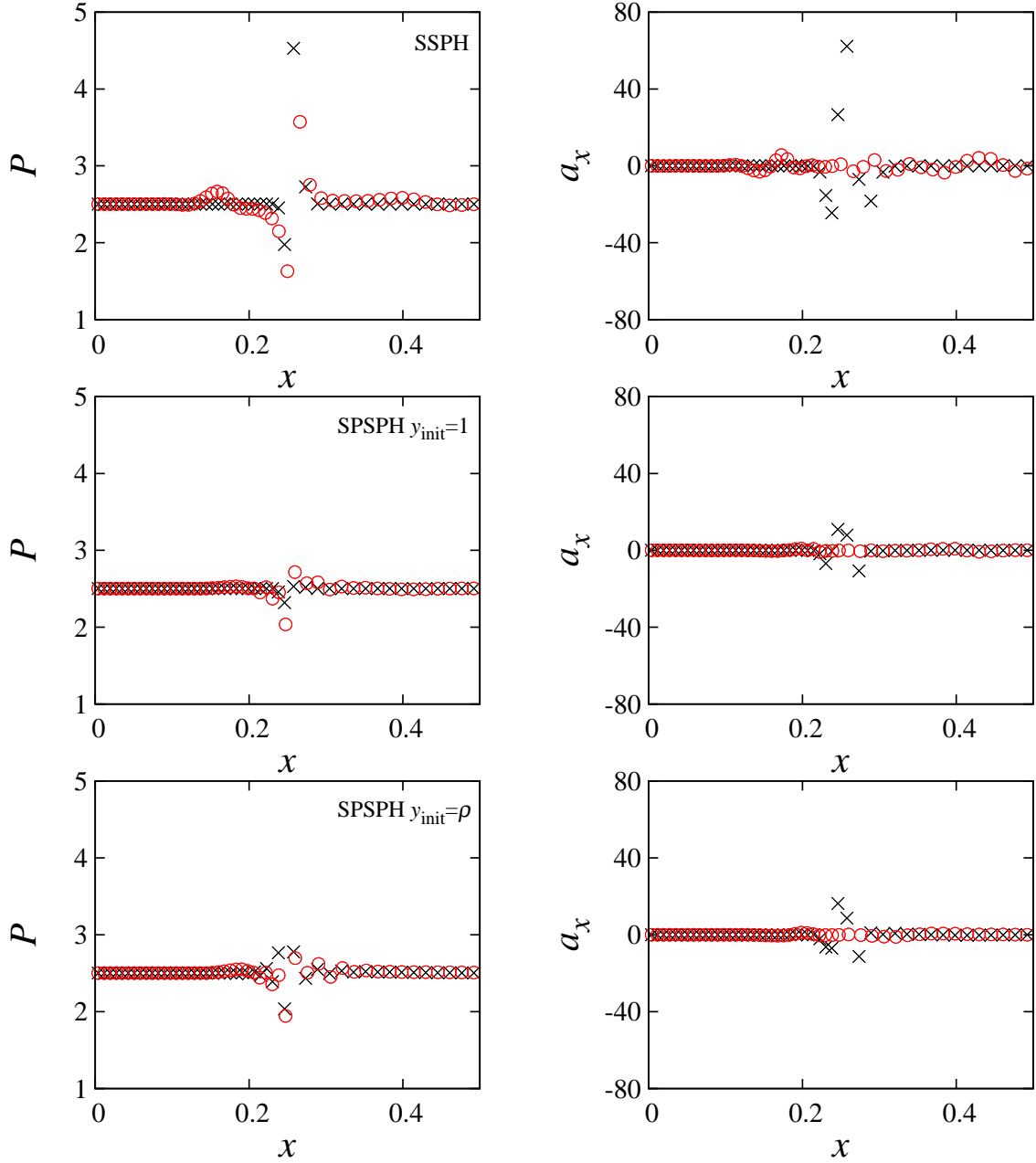}
\end{center}
\caption{Pressure (left) and acceleration (right) in the $x$ direction for three different models (run1,2 and 3).
Physical quantities which are only in the region of $|y|< 0.05$ at $t=0$ (crosses) and $t=0.1$ (circles).}
\end{figure}

\begin{figure}[htbp]
\begin{center}
\FigureFile(160mm,50mm){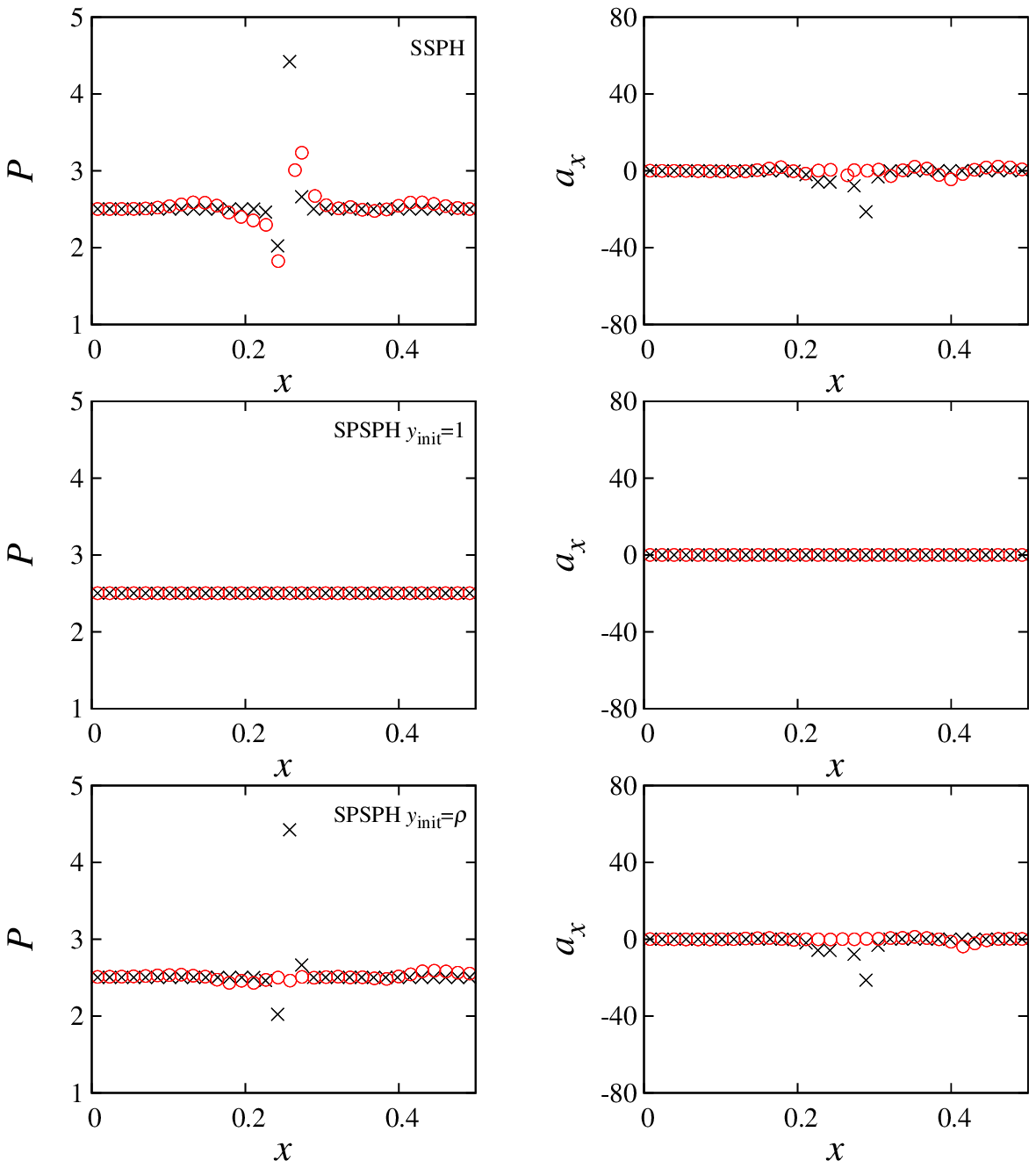}
\end{center}
\caption{The same as Fig. 4 but for runs 4 to 6 and with the mass ratio of 4:1.}
\end{figure}

\begin{figure}[htbp]
\begin{center}
\FigureFile(160mm,50mm){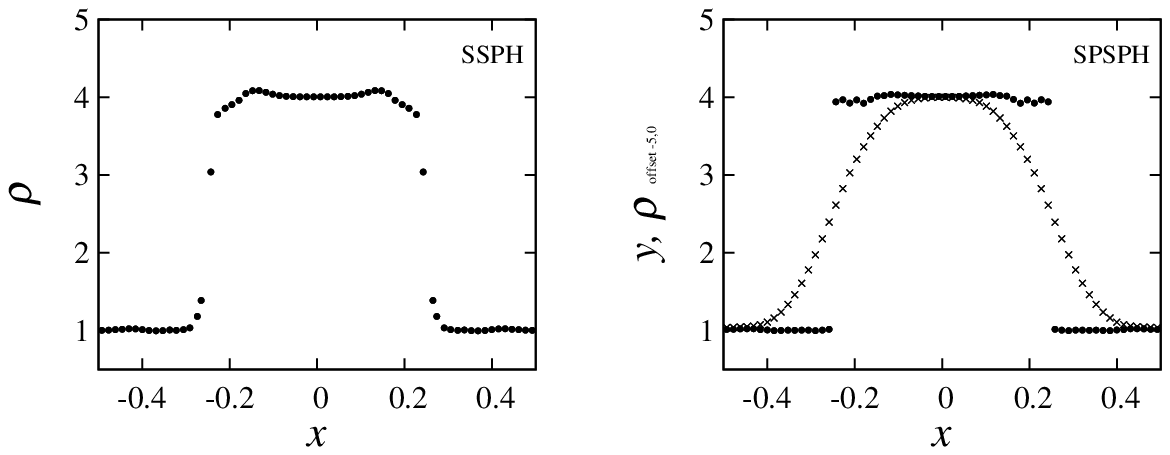}
\end{center}
\caption{The cross-section of the density and pseudo density. 
Values of particles in the region of $|y|< 0.05$ and plotted. 
The time $t=0.1$. 
The left hand-side panel shows the results of run 4 in table 1 (SSPH), and the right hand-side panel shows that of run 6 in table 1 (SPSPH). 
For the SPSPH run, we show both $y$ (crosses) and density (dots).}
\end{figure}
%%%%%%%%%%%%%%%%%%%%%%%%%%%%%%%%%%%%%%%%%%%%%%%%%%%%%%%%%%%%%%%%%%%%%%%%%%%%%%%%%%%%%%%%%%%%%
%
%Sodshock tube%%%%%%%%%%%%%%%%%%%%%%%%%%%%%%%%%%%%%%%%%%%%%%%%%%%%%%%%%%%%%%%%%%%%%%%%%%%%%%%

\subsection{Shock tube tests}
\subsubsection{Sod shock tube test}
The initial condition of one-dimensional shock tube test is the same as that used in Sod (1978). 
The computational domain is $-0.5\leq x < 0.5$ with a periodic boundary condition. 
We placed the discontinuity at the origin by setting initial condition: 
\begin{eqnarray}
\left\{
\begin{array}{l}
\rho=1\ \ \ \ \ \ \ P = 1\ \ \ \ \ \ \ \ \ \ \ v_x=0 \ \ \ \ \ \ \ {x<0},\\
\rho=0.25 \ \ P = 0.1795\ \ \ v_x=0 \ \ \ \ \ \ {x>0}.
\end{array}
\right.
\end{eqnarray}

We actually used a ``smoothed'' initial condition. 
The smoothed initial pressure is given by
\begin{eqnarray}
\left\{
\begin{array}{ll}
P = 1 & {x<-\frac{1}{C}},\\
P = \frac{P_l-P_h}{4}[(Cx)^3-3Cx]+\frac{P_h+P_l}{2} & {-\frac{1}{C}<x<\frac{1}{C}},\\
P = 0.1795 & {x>\frac{1}{C}}.
\end{array}
\right.
\end{eqnarray}
The coefficient $C$ is an arbitrary constant and we used 103.22 in this paper. 
$P_h$ and $P_l$ are pressures in the high and low density regions (i.e., 1 and 0.1795), respectively.
The smoothed initial density distribution is given by 

\begin{eqnarray}
\left\{
\begin{array}{ll}
\rho = 1 & {x<-\frac{1}{C}},\\
\rho = \frac{\rho_l-\rho_h}{4}[(C x)^3-3Cx]+\frac{\rho_h+\rho_l}{2} & {-\frac{1}{C}<x<\frac{1}{C}},\\
\rho = 0.25 & {x>\frac{1}{C}}.
\end{array}
\right.
\end{eqnarray}
where $\rho_h$ and $\rho_l$ are densities in high and low-density regions (i.e., 1 and 0.25), respectively. 

The position of the particle $a$, $x_a$ satisfies
\begin{eqnarray}
\int^{x_a}_{x_{a-1}} \rho(x) dx = m,
\end{eqnarray}
where, $m$ is the mass of a particle, and we used $m=1/1600$. 
We carried out two cases, i.e. $y_{\rm init}=1$ and $y_{\rm init}=\rho$. 
The initial pseudo mass is given by $Z = my/\rho = m/\rho$ for $y_{\rm init}=1$, and $Z=m$ for $y_{\rm init}=\rho$.
We set $D_{\rm dif} = D_M$ in this test.

%%%%結果%%%%
In Figure 7 we can see that the behaviors of SSPH and SPSPH calculations are very similar,
even though the volume elements used are quite different.
Thus, we can conclude that the use of pseudo density with artificial diffusion 
does not affect the behavior of SPH scheme for the standard shock tube test. 
In this case, the amplitude of the pressure wiggles of SPSPH with $y_{\rm init}=1$ are comparable with or smaller than those of SSPH.
In other words, SPSPH is not much better than SSPH in this case. 
The reason is that both SPSPH and SSPH have the error caused by the discontinuity of particle spacing.
We explain this further in section 4.5.
If we choose an adequate $y_{\rm init}$, the absolute value of the jump in pressure can be reduced to about half of that obtained in the SSPH run. This is because the distribution of the fundamental quantity $y$ in the SPSPH run with $y_{\rm init}$$ = \rho$ is much smoother than that with $y_{\rm init} = 1$ (see figure 8).

\begin{figure}[htbp]
\begin{center}
\FigureFile(160mm,50mm){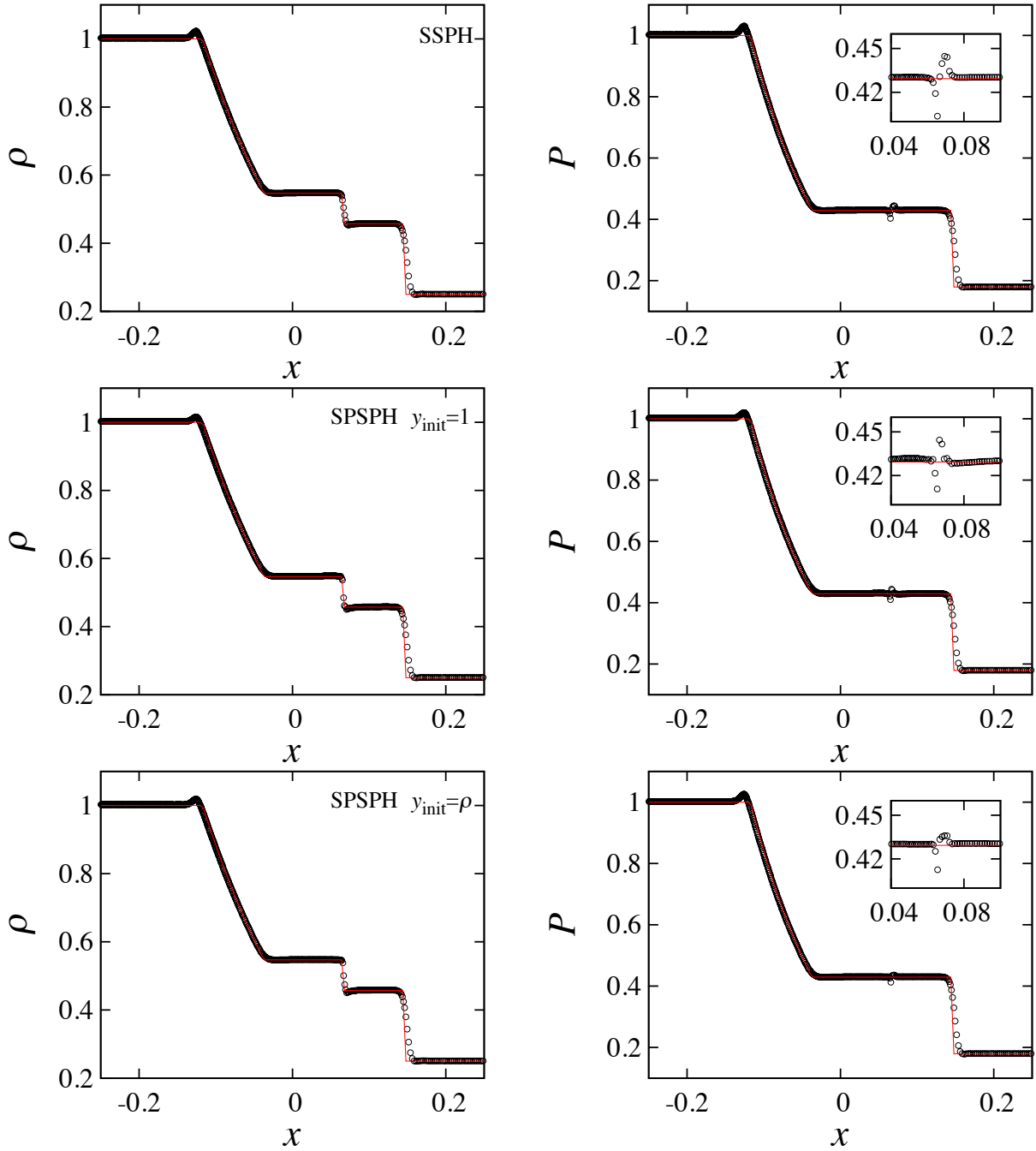}
\end{center}
\caption{Result of the sod shock test at $t=0.1$. 
Panels in top, middle, and bottom rows show the results of SSPH, SPSPH with $y_{\rm init}=1$, that with $y_{\rm init}=\rho$, respectively. 
The left hand-side panels show the density and the right side-hand panels show the pressure. 
Black circles show the numerical results, and red curves analytic solutions.
}
\end{figure}

\begin{figure}[htbp]
\begin{center}
\FigureFile(160mm,50mm){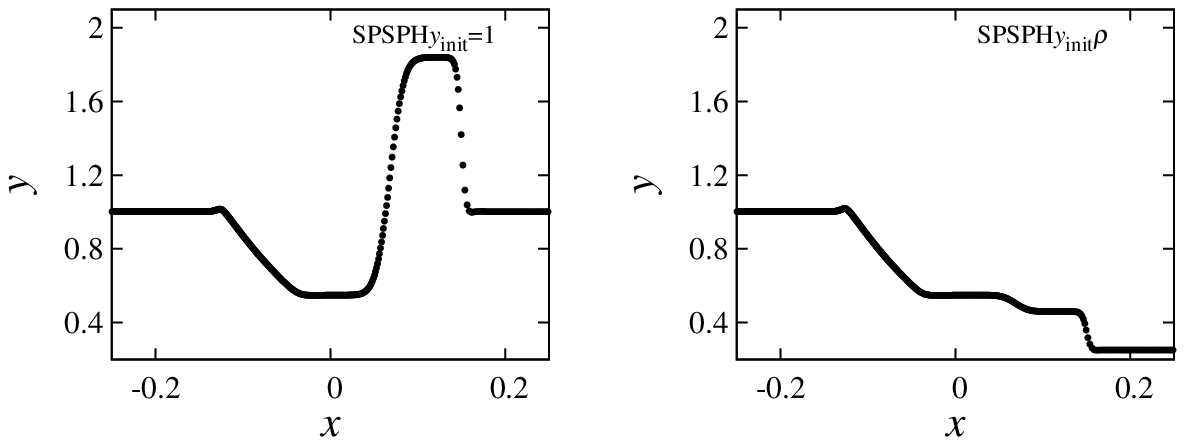}
\end{center}
\caption{The distribution of the pseudo density $y$ for the Sod shock tube test.
The left and right panels show the results of SPSPH with $y_{\rm init}=1$, 
and with $y_{\rm init}=\rho$, respectively.}
\end{figure}
%%%%%%%%%%%%%%%%%%%%%%%%%%%%%%%%%%%%%%%%%%%%%%%%%%%%%%%%%%%%%%%%%%%%%%%%%%%%%%%%%%%%%%%%%%%%
%
%strong-shock tube%%%%%%%%%%%%%%%%%%%%%%%%%%%%%%%%%%%%%%%%%%%%%%%%%%%%%%%%%%%%%%%%%%%%%%%%%%
\subsubsection{Strong shock tube test}
Here we present the result of a strong shock test, similar to what is performed in 
Toro (2009).
The computational domain is $-1.0\leq x < 1.0$ with a periodic boundary condition. 
We placed the discontinuity at the origin by setting initial conditions as 
\begin{eqnarray}
\left\{
\begin{array}{l}
\rho=1\ \ P = 1000.0\ \ \ v_x = 0 \ \ {x<0},\\
\rho=1 \ \ P = 0.01\ \ \ \ \ \ v_x = 0 \ \ {x>0}.
\end{array}
\right.
\end{eqnarray}
We used $\alpha_{\rm AV}=2$ and the same smoothed distribution as in Eq.(96) with $C = 129.025$, $y_{\rm init}=1$ everywhere, and $m=1/500$. 
We set $D_{\rm dif}=D_{\rm M}$, however $M$ equals $7$ with SPSPH. 

%%%%%結果%%%%%%%
In Figure 9, again, SSPH and SPSPH behave similarly. 
There are some differences. 
For example, the overshoot in the density at the contact discontinuity is 
smaller for SSPH, while the jump in the pressure is smaller for SPSPH. 
For shock tube tests, there is no reason to expect big improvement over SSPH. 
%%%%%%%%%%%%%%%%
%
\begin{figure}[htbp]
\begin{center}
\FigureFile(160mm,50mm){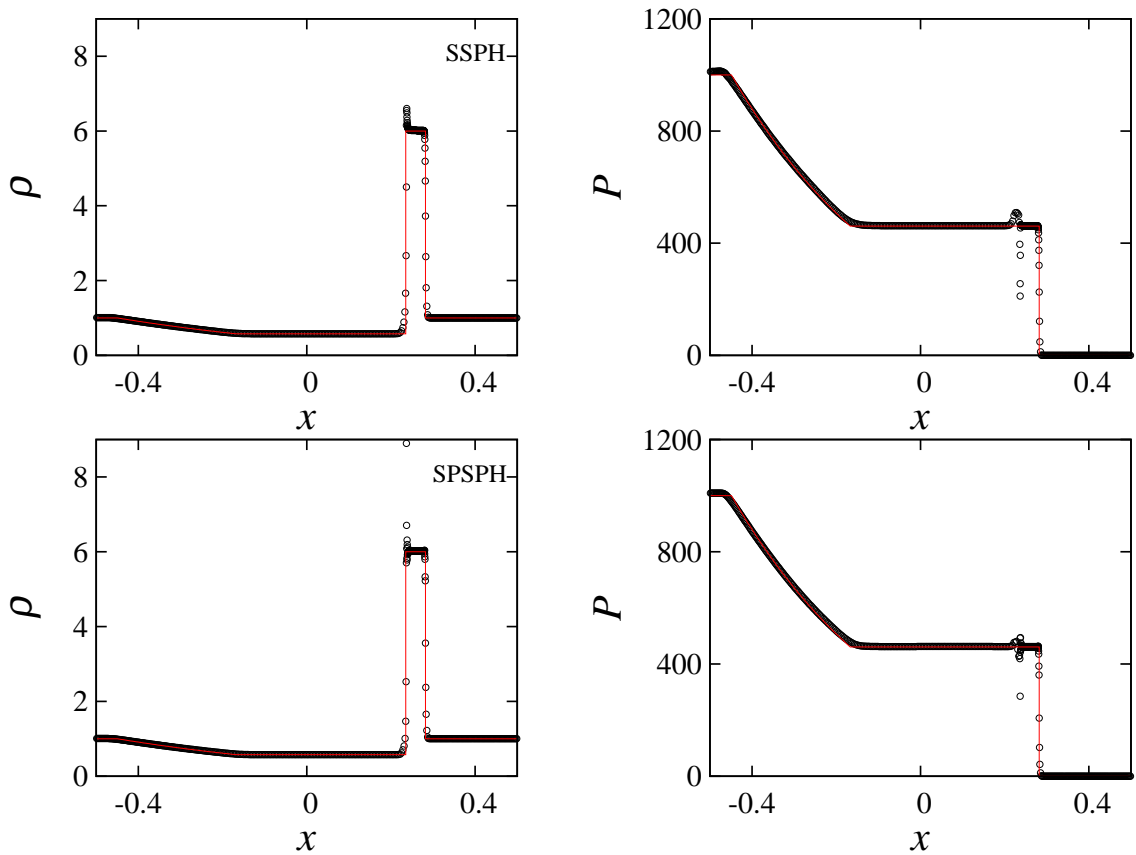}
\end{center}
\caption{Results of the one dimensional strong shock tube tests at $t=0.012$. 
Panels in the top and bottom sides show the results of SSPH and SPSPH runs, 
and left and right panels show the density and pressure. 
Black circles show the numerical results, and red curbs analytic solutions. }
\end{figure}
%
%%%%%%%%%%%%%%%%%%%%%%%%%%%%%%%%%%%%%%%%%%%%%%%%%%%%%%%%%%%%%%%%%%%%%%%%%%%%%%%%%%%%%%%%%%%%%%
%%
%%%KHI%%%%%%%%%%%%%%%%%%%%%%%%%%%%%%%%%%%%%%%%%%%%%%%%%%%%%%%%%%%%%%%%%%%%%%%%%%%%%%%%%%%%%%%%

\subsection{Kelvin-Helmholtz instability tests}
The KHI test is useful to investigate the ability of 
SPH schemes to handle hydrodynamical instabilities. (e.g., Okamoto et al. 2003, Agertz et al. 2007, Price 2008)

We performed two-dimensional calculations and use the computational domain that is a square of an unit size, 
$-0.5\leq x < 0.5$ and $-0.5\leq z < 0.5$, with a periodic boundary condition. 
We make the contact discontinuity by setting initial conditions as
\begin{eqnarray}
\left\{
\begin{array}{l}
\rho=2 \ \ -0.25\leq z\leq 0.25,\  \\
\rho=1 \ \ {\rm otherwise}.
\end{array}
\right.
\end{eqnarray} 
We set $P=2.5$, $\gamma = 5/3$ and ${v}_x(\equiv{v}_{x,h}) = 0.5$ in the dense region ,${v}_x(\equiv{v}_{x,l}) = -0.5$ in another. 
The initial velosity perturbation is
\begin{eqnarray}
{v}_z = -\frac{z}{|z|}A\sin\left\{\frac{2.0\pi(x+0.5)}{\lambda}\right\},\ \ (0.225 < |z| < 0.275) 
\end{eqnarray}
where $\lambda = 1/6$ and $A = 0.025$. The growth timescale of the KHI is
\begin{eqnarray}
\tau_{\rm KH}= \frac{\lambda(\rho_{h}+\rho_{l})}{\sqrt{\rho_h\rho_l}|{v}_{x,h}-{v}_{x,l}|}. 
\end{eqnarray}
For our setup, $\tau_{\rm KH}\simeq0.35$. This setup is the same as that used in Price (2008).

We used two different initial distributions for $y$. 
In the first one, the $y_{\rm init}$ is unity everywhere. 
In the second one, $y_{\rm init}$ = $\rho$. 
In this way, we can see if our scheme can handle the initial discontinuity of $y$.
For this run, we used $D_{\rm dif} = D_{\rm M}$ and $\Delta t_{\rm dif}=\Delta t_{\rm CFL}$. 
Unlike in the square test, we did not let pseudo density diffuse before we started the calculation.
Particles in the two regions have the same mass. 
Thus the number density of particle is different in the two regions,
262144 in the dense square, and 131072 in the other region. 

Figure 10 shows the time evolution up to $t = 8\tau_{\rm KH}$. 
It is clear that SPSPH is much better than SSPH in dealing with KHI. 
With SSPH, the perturbation grow but the roll-like structure characteristic of the KHI is suppressed. 
Moreover rolls break apart by $t=4-8\tau_{\rm KH}$. 
These are due to the effect of the artificial surface tension at the boundary of two fluids. 
In two SPSPH runs, the KHI grows well, and there is no effect of the artificial surface tension. 
SPSPH can handle hydrodynamical instability even if $y$ is initially discontinuous. 

%%
%%%密度比1:100の図開始%%%
\begin{figure}[htbp]
\begin{center}
\FigureFile(160mm,50mm,angle=90){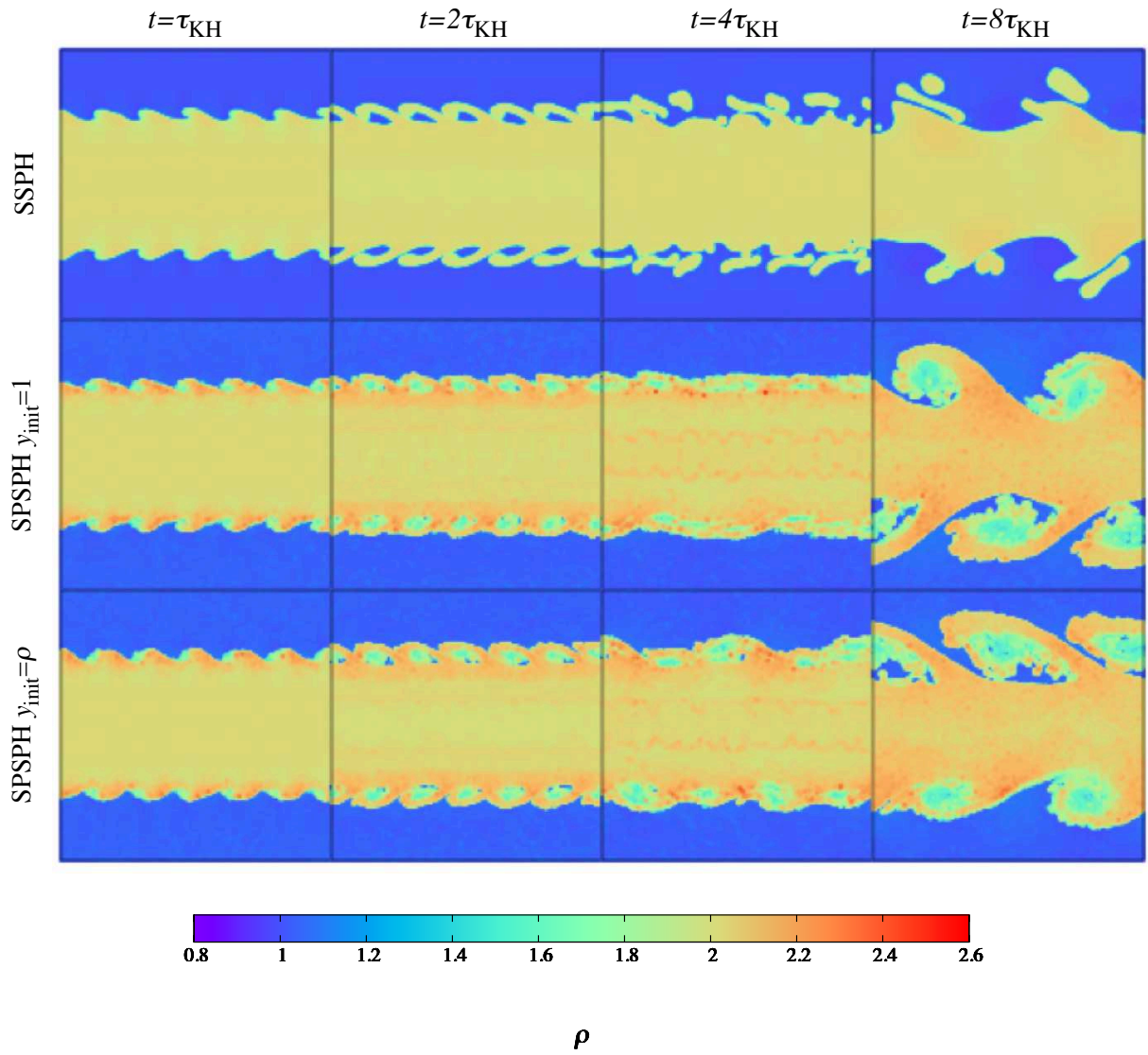}
\end{center}
\caption{Results of the KHI tests with the density ratio of 2:1. 
Density distributions at $t=1,2,4$ and $8$ are shown from left to right.
Panels in the top, middle, and bottom rows show the results of SSPH, SPSPH with $y_{\rm init}=1$, 
that with $y_{\rm init}=\rho$, respectively.}
\end{figure}
%%%%%%%%%%%%%%%%%%%%%%%%%%%%%%%%%%%%%%%%%%%%%%%%%%%%%%%%%%%%%%%%%%%%%%%%%%%%%%%%%%%%%%%%%%%%%
%
%きつめの静水圧平衡%%%%%%%%%%%%%%%%%%%%%%%%%%%%%%%%%%%%%%%%%%%%%%%%%%%%%%%%%%%%%%%%%%%%%%%%%%

\subsection{The square test with extreme density difference}

Here we present the results of the square test as in section 4.1, 
but with a much larger density construct. 
The density of the high-density region is 100 instead of four. 
We only did the cases with different mass particles since with equal-mass particles,  
the difference in the particle number density would become too large. 
For the SPSPH run with $y_{\rm init}=\rho$, we let $y$ evolve for 0.01 time unit before we start time integration.  
Figures 11 and 12 show the results, 
we can see that SPSPH can deal with very large density contrast much difficulty, 
even when the pseudo density is not initially continuous. 
%
%%%密度比1:100の図開始
\begin{figure}[htbp]
\begin{center}
\FigureFile(160mm,50mm){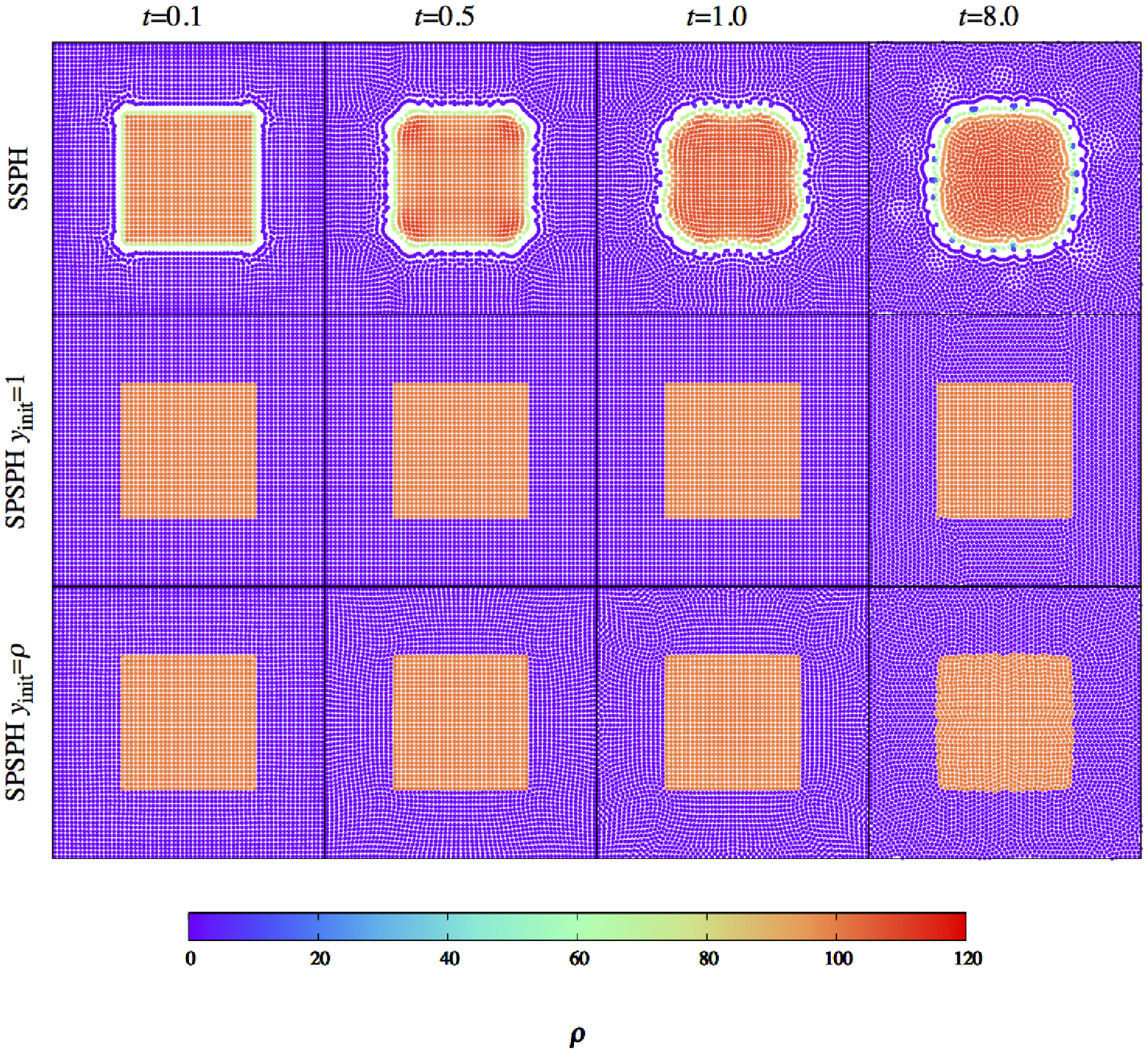}
\end{center}
\caption{Results of the square tests with the density ratio of 100:1. 
Density distributions at $t=0.1,0.5,1.0$ and $8.0$ are shown form left to right.
Panels in top, middle, and bottom rows show the results of SSPH, SPSPH with $y_{\rm init}$=1, 
that with $y_{\rm init}=\rho$, respectively.}
\end{figure}
%%%密度比1:100の図終了
\begin{figure}[htbp]
\begin{center}
\FigureFile(160mm,50mm){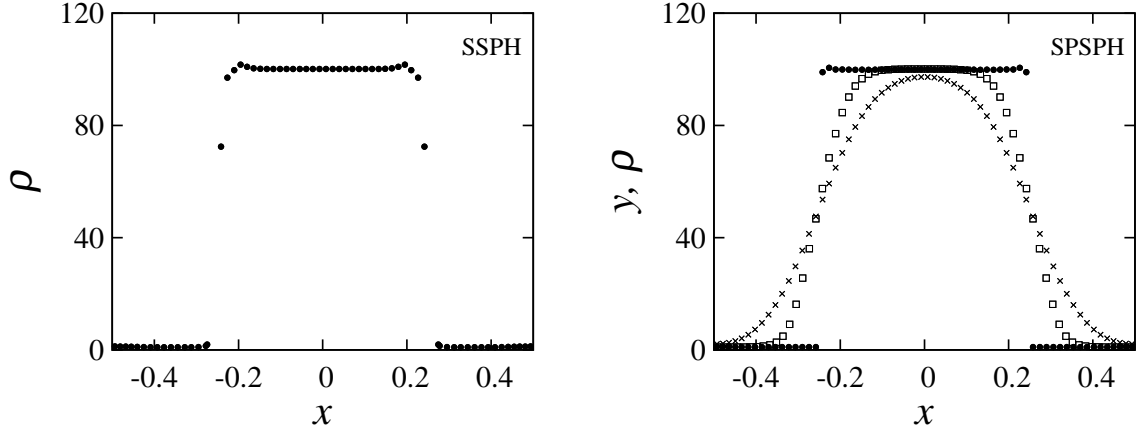}
\end{center}
\caption{The same at Fig. 6 but for the density contrast 100:1. 
In the right-hand-side panel, 
squares, crosses and dots show the profile of $y(t=0)$, $y(t=0.1)$ and density$(t=0.1)$, respectively.}
\end{figure}

%%%1dim静水圧平衡test%%%
\subsection{The one-dimensional hydrostatic equilibrium test}

Here, we investigate the ability of SSPH, DISPH and SPSPH to handle the contact discontinuity which exists in a hydrostatic equilibrium state with a discontinuous particle spacing. In this test, we consider the one-dimensional computational domain $-0.5\leq x<0.5$ with a periodic boundary condition. This computational domain is filled with a gas of $\gamma =1.4$. The initial condition is as follows:
\begin{eqnarray}
\left\{
\begin{array}{l}
\rho=1 \ \ P = 2.5\ \ v_x = 0 \ \ {x<0},\\
\rho=0.25 \ \ P = 2.5\ \ v_x = 0 \ \ {x>0}.fhn
\end{array}
\right.
\end{eqnarray}
The number of particles is 100, and mass of each particle is 1/160. Thus, the interparticle distance is 1/160 for
$x<0$ and 1/40 for $x>0$. The initial value of pseudo density $y_{\rm init}$ is set to unity and that of $Z$ in $x< 0$ is 1/160 and that in $x>0$ is 1/40.
The diffusion coefficient for SPSPH is set to $D_{\rm dif}=D_{M}$.

In the left panels of figure 13, we show the distributions of pressure and acceleration for the one-dimensional test for SSPH, DISPH, and SPSPH ($t = 0$).
We can see that the distributions of pressure and acceleration for SPSPH are identical to those for DISPH.
However these for SSPH are different. 
Although the wiggles of acceleration for SSPH is comparable to these for SPSPH and DISPH, that of pressure for SSPH is more than three times larger than those for SPSPH and DISPH.
Note that, of course, these quantities should be uniform throughout the computational domain.
Hence, these wiggles are induced by the asymmetric distribution of particles.\\
With these results, we can now understand the results show in section 4.1 and 4.2 better.
In the one-dimensional shock tube tests, we see the weak pressure wiggles at the contact discontinuities.
Thus wiggles are caused by the inhomogeneous particle distribution.
Even in DISPH, we observed these weak wiggles (see figure 1 in Saitoh \& Makino 2013) at the contact discontinuity.
In the panels in the right-hand side of figure 13, we show the result
of the same hydrostatic test as in the left-hand side panels, but in
two-dimensional calculation. The density contrast is 1:4 in both
cases, but the ratio of interparticle separation is 1:4 in 1D and 1:2
in 2D. Thus, we can expect that the effect of this inhomogeneity in
the particle distribution would be smaller for 2D (or even so for 3D)
calculations than in 1D calculation, even when the density contrast is
the same. Indeed, the wiggles in both pressure and acceleration are
much smaller in 2D than in 1D, for DISPH and SPSPH. In the case of
SSPH, the wiggle of the pressure is not much different for 1D and 2D
calculations. The error in the acceleration is somewhat smaller in 2D,
but this effect is not so drastic as in the case of DISPH and
SPSPH. Thus, we can expect that the improvement of SPSPH over DISPH is
quite large, for realistic, multi-dimensional calculations.

\begin{figure}[htbp]
\begin{center}
\FigureFile(160mm,50mm){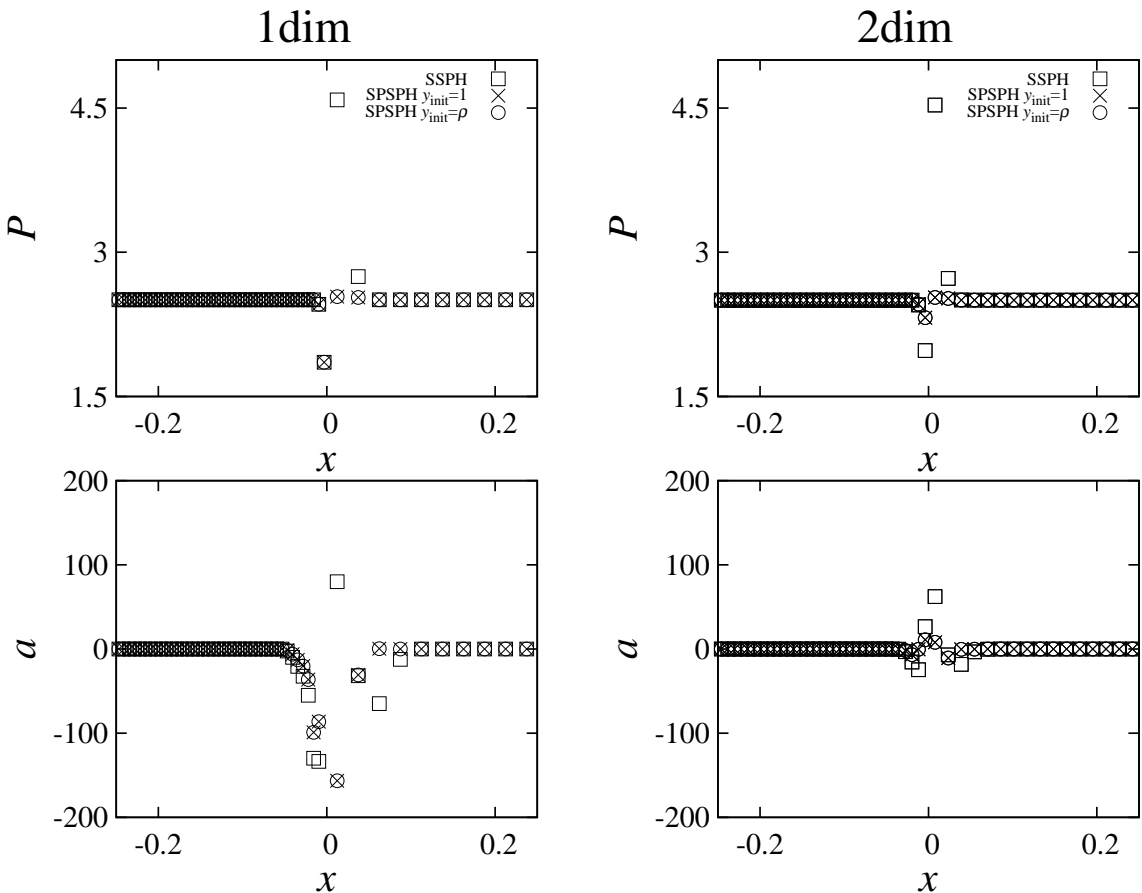}
\end{center}
\caption{Results of the one (left) and two (right) dimensional hydrostatic equilibrium test at t = 0.}
\end{figure}

%%%%%%%%%%%%%%%%%%%%%%%%%%%%%%%%%%%%%%%%%%%%%%%%%%%%%%%%%%%%%%%%%%%%%%%%%%%%%%%%%%%%%%%%%%%%

%議論%%%%%%%%%%%%%%%%%%%%%%%%%%%%%%%%%%%%%%%%%%%%%%%%%%%%%%%%%%%%%%%%%%%%%%%%%%%%%%%%%%%%%%%
\section{Discussion}
\subsection{Limiting cases for the diffusion constant}
As we discussed in section 2.2, if we set $D=0$, 
our SPSPH is reduced to either SSPH or the scheme proposed by Ott \& Schnetter (2003). 
Here, we consider the other limit of $D=\infty$. 
This means we would effectively solve an elliptic equation, 
instead of the parabolic diffusion equation. 
Therefore, the pseudo density $y$ would take the same value everywhere. 
If we set that value to unity, we have
\begin{eqnarray}
\sum_bZ_bW_{ab}(h_a)=1,
\end{eqnarray}
from Eq.(46). 
We can see that, in Eq.(104), $Z_b$ is determined purely from the positions of particles. 
Thus, one might think it would give an even better way to formulate SPH equations. 
We have performed some experiments with this form, so far with little success. 
It is hard to determine the value of $Z_b$ in Eq.(104).

With a matrix $\vec{Q}_{ab} = W_{ab}$ and vectors $\vec{R}_a = Z_a,\ \vec{s}_a = 1$, we can rewrite Eq.(104) into,
\begin{eqnarray}
\vec{Q}\vec{R} = \vec{s}.
\end{eqnarray}
The condition for the existence of a unique solution is that the matrix $\vec{Q}$ is regular. 
In other words, all eigenvalues of $\vec{Q}$ should be nonzero. 
We introduce two assumptions for simplicity.
First, we consider the one dimensional case. 
Second, we assume a uniform particle separation $\Delta x$. 
With these assumptions, the eigenvectors $\vec{q}$ of $\vec{Q}$ are given by
\begin{eqnarray}
\vec{q}_a = e^{ikx_a},
\end{eqnarray}
where $k$ is a wavenumber and satisfies $0\leq k \leq 2\pi/\Delta x$.
Hence, the eigenvalues are  
\begin{eqnarray}
\sum_b W_{ab} e^{ik(x_b-x_a)}&=& \sum_b \frac{1}{Z_b}Z_bW_{ab} e^{ik(x_b-x_a)},\nonumber\\
&=&\frac{1}{Z^0}\fourier{W}.
\end{eqnarray}
Here, we used the fact that for a uniform particle separation the value of $Z_b$ is constant anywhere.
Eq.(107) means that the fourier transform of $W$ must be nonzero in the range $0\leq k \leq 2\pi/\Delta x$ in order to have an inverse matrix of $\vec{Q}$.
However Dehnen \& Aly (2012) showed that a fourier transform of some kernel functions become zero at ${}^{\exists}k$ with the uniform particle separation. 
Moreover, we are not sure that those of other kernels such as Wendland functions do not become zero at ${}^{\exists}k$ for inhomogeneous particle distributions.
Therefore SPH with functions Eq.(104) is less than successful.

%%%%%%%%%%%%%%%%%%%%%%%%%%%%%%%%%%%%%%%%%%%%%%%%%%%%%%%%%

\subsection{Free surface}
The SSPH scheme cannot handle the free surface well, since the density of particles near the surface is grossly underestimated. 
Here we discuss the possibility to extend our method to handle free surface. 
As a simple example, we consider the water surface (like that of sea surface).
From the physical point of view, the surface of water is not free, but covered by the atmosphere. 
In other words, it is simply the contact discontinuity of water and air. 
Thus, if we express air as well as water, by SPH particles, the surface of the water will be handled properly. 
This is however impossible with SSPH, since it cannot handle large density jumps. 
However, with our SPSPH, density jump would not cause problems. 
Thus, one solution to the treatment of the free surface is to introduce SPH particles that represent thin air. 
This scheme would work fine for engineering problems in which there actually is air. 
It might also works for problems in planetary science, like the giant-impact simulations.

%%%%%%%%%%%%%%%%%%%%%%%%%%%%%%%%%%%%%%%%%%%%%%%%%%%%%%%%%

\section{Summary}
The SSPH scheme cannot handle discontinuities in density.  The reason is 
that SSPH requires that density is positive and spatial continuity even at 
the discontinuities of density.
To solve this problem we introduce a new quantity “pseudo-density” and 
require the continuity of pseudo-density instead of that of the density. 
Pseudo-density evolve with artificial diffusion for guarantee the positive and 
spatial differentiable. 
SPSPH can handle the contact discontinuities quite well and has possibility
for handling the free surface with particles that represent thin air. 
%%%%%%%%%%%%%%%%%%%%%%%%%%%%%%%%%%%%%%%%%%%%%%%%%%%%%%%%%%%%%%%%%%%%%%%%%%%%%%%%%%%%%%%%%%%%

\bigskip
%謝辞aa\\
\section*{Acknowledgement}
We thank the anonymous referee for her/his insightful comments and suggestions.
This work is supported by MEXT SPIRE and JICFuS and JSPS Grants-in-Aid for Scientific Research (26707007).

%\bibliographystyle{/Users/yamamototomoko/desktop/apj}
%\bibliography{/Users/yamamototomoko/desktop/my}

%\end{document}
%\end{verbatim}
\end{document}